\begingroup%%%%%%%%%%%%%%%%%%%%%%%%%%%%%%%%%%%%%%%%%%%%%%%%%%%%%%%%%%%%%%%%%%%%%
\documentclass[journal=jpclcd,manuscript=letter]{achemso}
\usepackage[version=3]{mhchem} % Formula subscripts using \ce{}
\usepackage{placeins}
\usepackage{hyperref}
\usepackage{braket}

\usepackage[authormarkuptext=name,authormarkup=none,final]{changes}
\definecolor{acolour}{RGB}{0, 0, 255}
\definecolor{bcolour}{RGB}{255, 0, 0}
\definechangesauthor[name={Yorick}, color={acolour}]{YS}

\makeatletter
\newcommand*\rel@kern[1]{\kern#1\dimexpr\macc@kerna}
\newcommand*\widebar[1]{%
  \begingroup
  \def\mathaccent##1##2{%
    \rel@kern{0.8}%
    \overline{\rel@kern{-0.8}\macc@nucleus\rel@kern{0.2}}%
    \rel@kern{-0.2}%
  }%
  \macc@depth\@ne
  \let\math@bgroup\@empty \let\math@egroup\macc@set@skewchar
  \mathsurround\z@ \frozen@everymath{\mathgroup\macc@group\relax}%
  \macc@set@skewchar\relax
  \let\mathaccentV\macc@nested@a
  \macc@nested@a\relax111{#1}%
  \endgroup
}

\newcommand*\spinup[1]{\overline{#1}}

\author{Yorick L. A. Schmerwitz}
\affiliation[University of Iceland]
{Science Institute of the University of Iceland, VR-III, 107 Reykjav\'{\i}k, Iceland}

\author{Aleksei V. Ivanov}
\affiliation[University Iceland]
{Science Institute of the University of Iceland, VR-III, 107 Reykjav\'{\i}k, Iceland}

\author{Elvar \"O. J\'{o}nsson}
\affiliation[University Iceland]
{Science Institute of the University of Iceland, VR-III, 107 Reykjav\'{\i}k, Iceland}

\author{Hannes Jónsson}
\affiliation[University of Iceland]
{Faculty of Physical Sciences, University of Iceland, VR-III, 107 Reykjav\'{\i}k, Iceland}
\alsoaffiliation{Department of Applied Physics, Aalto University, FI-00076 Espoo, Finland}

\author{Gianluca Levi}
\affiliation[University of Iceland]
{Science Institute of the University of Iceland, VR-III, 107 Reykjav\'{\i}k, Iceland}
\email{giale@hi.is,hj@hi.is}

\title{Variational Density Functional Calculations of Excited States: Conical Intersection and Avoided Crossing in Ethylene Bond Twisting}

\begin{document}

\renewcommand*\tocentryname{TOC Graphic}
\begin{tocentry}
   \includegraphics[width = 0.99\textwidth]{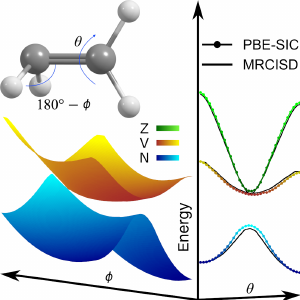}
\end{tocentry}

\begin{abstract}
Theoretical studies of photochemical processes require a description of the energy surfaces of excited electronic states, especially near degeneracies, where transitions between states are most likely. Systems relevant to photochemical applications are typically too large for high-level multireference methods, and while time-dependent density functional theory (TDDFT) is efficient, it can fail to provide the required accuracy. A variational, time-independent density functional approach is applied to the twisting of the double bond and pyramidal distortion in ethylene, the quintessential model for photochemical studies. By allowing for symmetry breaking, the calculated energy surfaces exhibit the correct topology around the twisted-pyramidalized conical intersection even when using a semilocal functional approximation, and by including explicit self-interaction correction, the torsional energy curves are in close  agreement with published multireference results. The findings of the present work point to the possibility of using a single determinant time-independent density functional approach to simulate nonadiabatic dynamics, even for large systems where multireference methods are impractical and TDDFT is often not accurate enough.
\end{abstract}

%%%%%%%%%%%%%%%%%%%%%%%%%%%%%%%%%%%%%%%%%%%%%%%%%%%%%%%%%%%%%%%%%%%%%

%%% Introduction
Configurations of atoms where different electronic states are close in energy play a key role in photoreactions. For example, in photoinduced bond-breaking processes, the bonding and antibonding frontier molecular orbitals tend to become degenerate, thereby facilitating dissociation into molecular fragments. Nuclear configurations where electronic states are degenerate, so-called conical intersections (CoIns)\cite{ConicalIntersectionsBook2011, ConicalIntersectionsBook2004, Yarkony2001}, promote nonadiabatic electronic transitions, as has been demonstrated theoretically\cite{Matsika2021, Malhado2014, Matsika2011, Levine2007, Worth2004, Teller1937} and experimentally\cite{wolf2019, Kowalewski2015, Polli2010}. The modeling of excited state energy surfaces near degeneracies is, therefore, fundamental for gaining an understanding of photochemical reactivity. It is an important challenge to carry out simulations of the dynamics of the atoms of large and complex systems in excited electronic states and predict the extent to which electronic transitions can occur.  

Near-degenerate electronic states are inherently multiconfigurational and have significant static electron correlation. Within wave function theory, this type of correlation can be accurately described for both ground and excited states\cite{Lischka2018} by including a few electron configurations in multireference configuration interaction (MRCI)\cite{Eter2012} or multiconfigurational self-consistent field (MCSCF)\cite{Eter2012, knowles1985efficient, werner1985second} calculations. Dynamic correlation, on the other hand, is described by including a large number of configurations, or through perturbation theory. In any case, however, the computational effort is large and it increases rapidly as the number of electrons increases. Kohn-Sham\cite{KohnSham1965} (KS) density functional theory\cite{hohenberg1964density} (DFT) treats dynamic correlation in a way that strikes a practical balance between accuracy and computational cost. However, approximate exchange-correlation functionals often fail to capture static correlation since DFT is a single Slater determinant formalism\cite{Yu2016, Cohen2008, Cremer2002, Cremer2001, Savin1996}. This deficiency also affects the commonly used linear-response time-dependent DFT\cite{runge1984density} approach (in the following referred to as TDDFT), where ground state orbitals and functionals (within the so-called adiabatic approximation) are used to estimate excited states. This approach leads, in particular, to inaccuracy near CoIns. Due to the lack of doubly excited configurations, the dimensionality of the seam between ground and excited states in a TDDFT calculation is $M$-1\cite{levine2006conical}, where $M$ is the number of internal nuclear degrees of freedom, whereas it should be $M$-2\cite{Matsika2011}. Other detrimental features of TDDFT are discontinuities in the relative energies caused by a sudden change of the reference state\cite{Barca2018}, and incorrect ordering of the energy of the electronic states, causing negative energy gaps\cite{barbatti2014surface}.

Several modifications of the standard TDDFT approach have been proposed to improve the description of excited states near degeneracies. The spin-flip TDDFT\cite{Shao2003} has proven successful in the description of bond dissociation energy curves\cite{wang2004time,Shao2003} and CoIns\cite{Winslow2020}. The dual-functional Tamm-Dancoff approximation\cite{shu2017dual, Shu2017}, and the more recent TDDFT configuration interaction method with one doubly excited configuration\cite{Athavale2021, Teh2019} have been specifically designed to recover the correct dimensionality of CoIns. \added[id=YS]{The particle-particle\cite{Yang2016} and hole-hole\cite{Bannwarth2020} Tamm-Dancoff approximations also yield correct CoIn topologies for the subsets of electronic states that they can describe (excitations from the HOMO for the former and excitations to the LUMO for the latter).} Extensions of the time-independent KS-DFT approach have been presented where an explicit multiconfigurational treatment of correlation\cite{Lischka2018,Ghosh2018} is included, as for example semiempirical DFT-MRCI\cite{grimme1999combination}, constrained DFT-CI\cite{Wu2007}, CAS-DFT approaches\cite{gritsenko2018efficient, nakata2006casscf, Grafenstein2005}, multiconfigurational pair-DFT (MC-PDFT)\cite{li2014multiconfiguration}, and spin-restricted ensemble-referenced KS (REKS)\cite{filatov2017description, Filatov1999}. The introduction of a multiconfigurational wave function, however, increases the computational cost significantly, and analytical atomic forces and nonadiabatic coupling between electronic states are, moreover, often not available, preventing the use of such approaches in simulations of the excited state atomic dynamics. \added{One exception is the state-interaction state-averaged REKS method with a small active space of two electrons in two orbitals, which has been used to describe systems with as many as 70 atoms in multiscale nonadiabatic dynamics simulations\cite{Liang2021,Yu2019} using analytical gradients and nonadiabatic couplings.\cite{Liu2021}}

Static correlation effects can in some cases be described by breaking the symmetry of a single determinant, as the broken-symmetry determinants can have an implicit multiconfigurational character\cite{Perdew2021,Yu2016,Cohen2008,Grafenstein2002,Cremer2002,Cremer2001,Grafenstein2000,Wittbrodt1996}. Broken-spin-symmetry solutions emerge in bond-breaking processes at so-called Coulson-Fischer points\,\cite{coulson1949}, representing the onset of static correlation\cite{Jake2018,Toth2016,Jimenez-Hoyos2011,Li2009}. This emergence is, for example, well known in the case of Hartree-Fock or KS-DFT calculations of the stretching of the bond in the dihydrogen molecule. Beyond a Coulson-Fischer point, both broken-spin-symmetry and spin-restricted solutions exist. In the stretched H$_2$ molecule, the spin-restricted solution has too high energy and is characterized by a triplet instability\cite{Sharada2015,Li2009}. The broken-symmetry solution allows for the spatial distribution of the spin-up and spin-down electrons to be different and gives good estimate of the energy. Symmetry breaking further leads to adequate estimates of singlet-triplet energy splittings in diradicals\cite{Grafenstein2002,Grafenstein2000} and torsional energy barriers\cite{Shao2003}. Spin symmetry breaking is formally justified within the pair-density interpretation of DFT\cite{Perdew1995}.

Symmetry breaking is expected to be similarly important in calculations of excited states. Excited states can be found as higher-energy self-consistent field (SCF) solutions of the KS equations, corresponding to stationary points on the electronic energy surface other than the ground state minimum.\cite{Burton2022,Ivanov2021jctc,Levi2020jctc,Levi2020fd,Hait2021,Carter-Fenk2020,Hait2020,Barca2018,Ayers2015,Peng2013,Gilbert2008,Perdew1985} Unlike TDDFT, such calculations are variational (i.\ e.\ based on the calculus of variation), and can provide better approximations to long-range charge-transfer\cite{Hait2021,Barca2018,Zhekova2014}, Rydberg\cite{Seidu2015,Cheng2008}, core-level\cite{Besley2021,Besley2009}, and other excitations\cite{Hait2021,Hait2020} characterized by significant change in the electron density. However, while excited state DFT has been applied in a variety of fields\cite{NiklasGjerding2021,Daga2021,BourneWorster2021,Malis2021,Levi2020,Levi2018,hellman2004potential}, only a few studies of electronic near-degeneracies have been reported\cite{pradhan2018non,Barca2018,Ramos2018} and the results appear contradictory. According to Barca {\it et al.} \cite{Barca2018}, calculations using the BLYP\cite{PhysRevA.38.3098, PhysRevB.37.785} and B3LYP\cite{doi:10.1063/1.464913} functionals can successfully predict a CoIn in retinal, although only one branching space coordinate was considered. Ramos and Pavanello\cite{Ramos2018}, however, reported that calculations with B3LYP fail to describe a CoIn of H$_3$, and Pradhan {\it et al.}\cite{pradhan2018non} presented spurious crossings and negative energy gaps close to the twisted and pyramidalized CoIn of ethylene calculated using the PBE functional\cite{perdew1997generalized}. From the studies reported so far, it is not clear whether a single determinant variational DFT approach can describe phenomena such as avoided crossings and CoIns.  

We address this question here by analyzing the twisting and pyramidalization of the ethylene molecule. Ethylene is the quintessential model system for photoisomerization and nonadiabatic processes in organic compounds. It is commonly used as a test system for multiconfigurational and nonadiabatic approaches\cite{Malis2021,pradhan2018non,Lischka2018,Shu2017,Zhang2015,Tsuchimochi2015,barbatti2014surface}, and its ground and excited state energy surfaces have been extensively characterized using high-level multireference methods\cite{Barbatti2004,Ben-Nun2000}. Twisting of the C=C double bond gives rise to an avoided crossing between the singlet ground state and the lowest doubly excited singlet state. At a torsional angle of 90$^\circ$, the former acquires diradical character while the latter becomes ionic\cite{Salem1972}. Double bond twisting and distortion of a methylene group towards a pyramidal shape lead to a CoIn between the ground and \added[id=YS]{a singly excited state}\cite{barbatti2014surface,Barbatti2004}. Hereafter, the singlet ground state will be denoted as N, the singly excited state as V and the doubly excited state as Z, following the Mulliken notation\cite{Merer1969}. 

As shown below, we find that time-independent density functional calculations that allow for broken symmetry can provide a good description of the energy surfaces of all three states, N, V and Z, even close to the CoIn, whereas the results of TDDFT calculations are known to have significant errors\cite{barbatti2014surface}. When multiple solutions coexist, the calculations can easily converge to solutions that preserve symmetries and give poor descriptions of the energy surfaces. This circumstance can explain the seemingly contradictory results obtained by Pradhan {\it et al.}\cite{pradhan2018non} using the same variational, time-independent DFT formalism. While semilocal, generalized gradient approximation functionals, such as PBE\cite{perdew1997generalized}, give energy surfaces of similar shape as those obtained from MRCISD calculations, there can be a large error on the energy gaps because the self-interaction error, which is inherent in practical implementations of KS-DFT, has different magnitude for the different electronic states. By applying explicit Perdew-Zunger self-interaction correction\cite{Perdew1981}, the calculated energy gaps are improved and become close to the MRCISD results.

%  Methods:

Calculations of electronic states near a degeneracy are challenging, because (1) standard SCF algorithms are prone to convergence failure when near-degenerate orbitals are unequally occupied\cite{Levi2020jctc,Levi2020fd,Voorhis2002}; (2) special care is needed when choosing the initial orbitals in order to obtain broken-symmetry solutions \cite{Vaucher2017,Filatov1999jcp}; and (3) excited states are typically saddle points on the electronic energy surface, making it necessary to estimate the degree(s) of freedom for which the energy needs to be maximized. Recently, we presented a direct orbital optimization method\cite{Ivanov2021jctc,Ivanov2021cpc,Levi2020jctc} based on an exponential transformation and quasi-Newton algorithms for finding both minima and saddle points of any order. This method is suited for calculations of electronically degenerate systems because direct optimization algorithms have proven more robust than traditional SCF methods in such cases\cite{Levi2020jctc,Levi2020fd,Voorhis2002}. It can also be used with functionals, such as the Perdew-Zunger self-interaction corrected (SIC) functional\cite{Perdew1981}, which include orbital density dependence\cite{Ivanov2021jctc} and, therefore, unlike KS functionals, are not unitary invariant. This direct optimization method is used here to calculate the energy of the N, V and Z states of ethylene as a function of the torsional and pyramidalization angles ($\theta$ and $\phi$, respectively). 

%  Results:

In the following, a Slater determinant is indicated by the electronic configuration including only the two highest occupied spin orbitals (frontier or open-shell orbitals), e.g. $\ket{ x \spinup x}$. All energies are evaluated for optimized determinants corresponding to stationary solutions. The symmetry-adapted multireference wave function for a given state is indicated with $\ket{\Psi}$ together with the dominant electronic configuration of the state as a subscript.

\begin{figure}
	\centering
	\includegraphics[width = 1\textwidth]{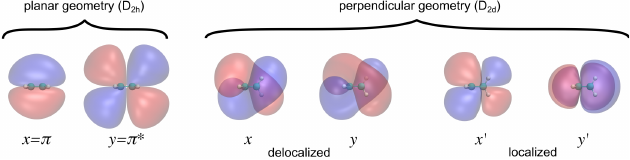}
	\caption{HOMO and LUMO orbitals of ethylene at the planar geometry and open-shell delocalized and localized orbitals at the perpendicular geometry. The orbitals are obtained from ground state PBE calculations and are rendered
for isovalues of $\pm 0.034$\,a.u.}
	\label{fig:fig1}
\end{figure}

At the planar geometry of ethylene ($\theta = \phi = 0^\circ$), the KS-DFT calculation for the singlet ground state, N, gives a closed-shell Slater determinant denoted as $\ket{ x \spinup x}$, where $x$ is the HOMO with character of a $\pi$ orbital. The HOMO is energetically well-separated from the LUMO, which is a $\pi^{*}$ orbital. The Z state is obtained by promotion of two electrons into the ground state LUMO and corresponds to a second-order saddle point on the electronic energy surface. The direct optimization converged to this saddle point gives a closed-shell Slater determinant denoted as $\ket{ y \spinup y}$. The $x$ and $y$ frontier orbitals from a PBE calculation at the ground state geometry are shown in Figure \ref{fig:fig1}.
% ----------------  figure 1 ----------------------
% -------------------------------------------------
As the torsional angle $\theta$ is increased, a Coulson-Fisher point appears at around $\theta = 60^\circ$ and two sets of N and Z solutions emerge. One set consists of the spin-restricted determinants $\ket{ x \spinup x}$ and $\ket{ y \spinup y}$, where $x$ and $y$ are delocalized orbitals that become degenerate at the perpendicular geometry ($\theta = 90^\circ$). The $x$ and $y$ orbitals at the perpendicular geometry are shown in Figure \ref{fig:fig1}. The determinants $\ket{ x \spinup x}$ and $\ket{ y \spinup y}$ converged around $\theta = 90^\circ$ correspond to first-order saddle points from a spin-unrestricted perspective. They are pure singlets and further preserve the spatial symmetry of the wave function (and electron density) in the D2 point group of twisted ethylene ($0^\circ < \theta < 90^\circ$). The second set of N and Z solutions consists of the determinants $\ket{ x^{\prime} \spinup y^{\prime}}$ and $\ket{ y^{\prime} \spinup y^{\prime}}$ (degenerate with $\ket{y^{\prime} \spinup x^{\prime}}$ and $\ket{ x^{\prime} \spinup x^{\prime}}$, respectively), where $x^{\prime}$ and $y^{\prime}$ are localized frontier orbitals. Initial orbitals for these solutions can be generated by taking a linear combination of the delocalized $x$ and $y$ orbitals:\cite{Grafenstein2002,Cremer2001,Grafenstein2000,szabo1996modern}
\begin{align}\label{eq:eq1}
    x^{\prime} & = \cos{(\gamma)}x + \sin{(\gamma)}y\,,\\
    \label{eq:eq2}
    y^{\prime} & = \cos{(\gamma)}x - \sin{(\gamma)}y
\end{align}
In eqs \ref{eq:eq1} and \ref{eq:eq2}, $\gamma$ can be between $0^\circ$ and $45^\circ$, but only $\gamma$ larger than 0 leads to the two additional solutions $\ket{ x^{\prime} \spinup y^{\prime}}$ and $\ket{ y^{\prime} \spinup y^{\prime}}$ beyond the Coulson-Fisher point. The $x^{\prime}$ and $y^{\prime}$ localized orbitals at the perpendicular geometry are shown in Figure \ref{fig:fig1}. The determinants $\ket{ x^{\prime} \spinup y^{\prime}}$ and $\ket{ y^{\prime} \spinup y^{\prime}}$ break the spin and spatial symmetry, respectively, compared to the multireference wave functions of the N and Z states ($\ket{^1\Psi_{x^2-y^2}}$ and $\ket{^1\Psi_{x^2+y^2}}$ at $\theta = 90^\circ$), as can be seen by expanding $\ket{ x^{\prime} \spinup y^{\prime}}$ and $\ket{ y^{\prime} \spinup y^{\prime}}$ using eqs\,\ref{eq:eq1} and \ref{eq:eq2}, while neglecting orbital relaxation: 
\begin{align}\label{eq:eq3}
    \ket{ x^{\prime} \spinup y^{\prime}} & = \cos^2{(\gamma)} \ket{ x \spinup x} - \sin^2{(\gamma)} \ket{ y \spinup y} -
                                            \cos{(\gamma)}\sin{(\gamma)}\left( \ket{ x \spinup y} - \ket{ y \spinup x} \right) \nonumber \\
                                        & = \cos^2{(\gamma)} \ket{ x \spinup x} - \sin^2{(\gamma)} \ket{ y \spinup y} -
                                            \cos{(\gamma)}\sin{(\gamma)} \ket{^3\Psi_{xy}}\,,\\
    \label{eq:eq4}
    \ket{ y^{\prime} \spinup y^{\prime}} & = \cos^2{(\gamma)} \ket{ x \spinup x} + \sin^2{(\gamma)} \ket{ y \spinup y} -
                                            \cos{(\gamma)}\sin{(\gamma)}\left( \ket{ x \spinup y} + \ket{ y \spinup x} \right) \nonumber \\
                                        & = \cos^2{(\gamma)} \ket{ x \spinup x} + \sin^2{(\gamma)} \ket{ y \spinup y} -
                                            \cos{(\gamma)}\sin{(\gamma)} \ket{^1\Psi_{xy}}\,.
\end{align}
Inclusion of the triplet $\ket{^3\Psi_{xy}}$ wave function in $\ket{ x^{\prime} \spinup y^{\prime}}$ breaks the spin symmetry of the N state, while inclusion of the open-shell singlet wave function $\ket{^1\Psi_{xy}}$  in $\ket{ y^{\prime} \spinup y^{\prime}}$ breaks the spatial symmetry of the Z state, because $\ket{^1\Psi_{xy}}$ transforms according to a different irreducible representation of the molecular point group than the multireference wave function of the Z state. The $\ket{ x^{\prime} \spinup y^{\prime}}$ and $\ket{ y^{\prime} \spinup y^{\prime}}$ solutions correspond to a minimum and to a second-order saddle point on the electronic energy surface, respectively.

The \added[id=YS]{singly} excited state, V, is an open-shell singlet and has the same electron configuration at all torsional angles. Its energy is approximated using the spin purification formula\cite{Ziegler1977}
\begin{align}\label{eq:eq5}
E_{\text{V}} = 
2E(\ket{x \spinup y})  - E(\ket{xy})\,.
\end{align}
The spin-mixed $\ket{x \spinup y}$ determinant is obtained with the direct optimization method by converging to a first-order saddle point on the electronic energy surface. The triplet $\ket{xy}$ determinant is found independently by minimization as it is the lowest-energy triplet state.

\begin{figure}
	\centering
	\includegraphics[width = \textwidth]{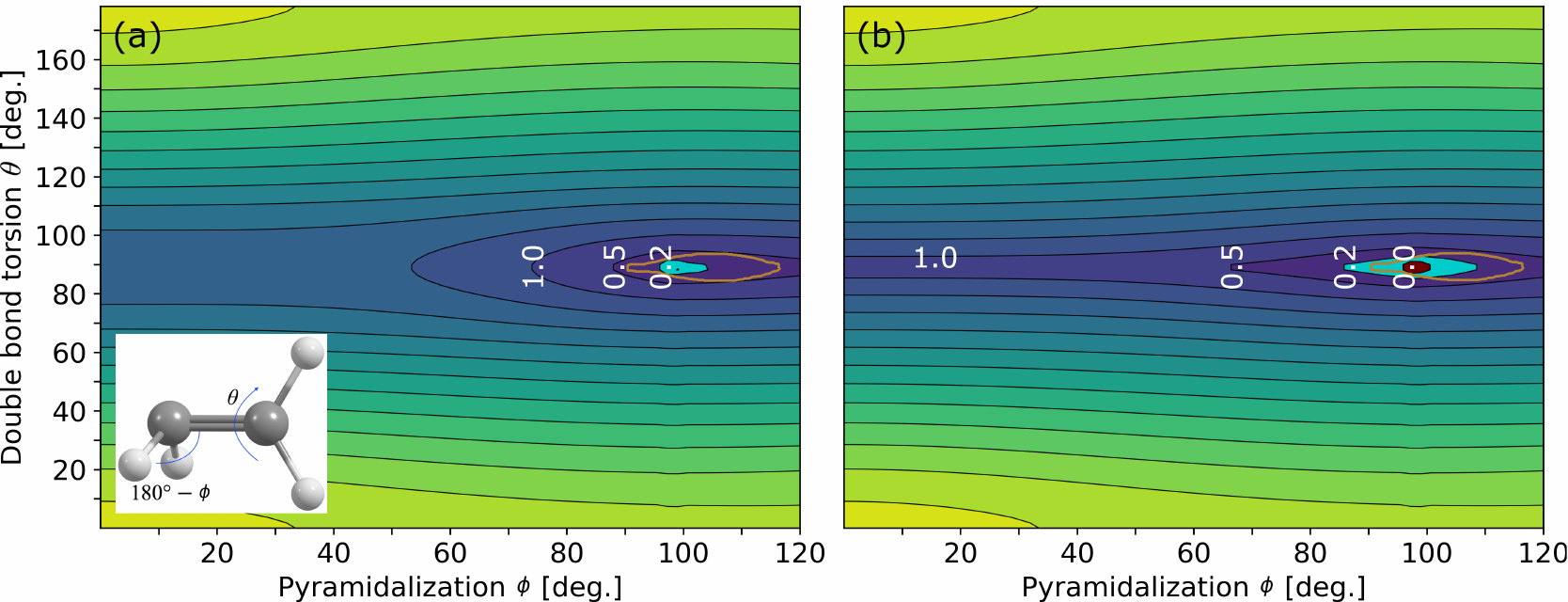}
	\caption{Energy gap \added[id=YS]{in eV} between the open-shell singly excited state, V, and the ground state, N, as a function of the torsional angle around the C=C bond, $\theta$, and pyramidal distortion of one of the two methylene groups, $\phi$. The molecular geometry for $\theta = 90^{\circ}$ and $\phi = 70^{\circ}$ is shown in the inset. (a)  The ground state is taken to be the spin-unrestricted, broken-symmetry solution $\ket{ x^{\prime} \spinup y^{\prime}}$. (b) The ground state is taken to be the spin-restricted solution $\ket{ x \spinup x}$. In both cases, the energy of the V state is estimated from calculations of $\ket{x \spinup y}$ and $\ket{xy}$ and the use of the spin purification formula, eq\,\ref{eq:eq5}. Regions colored in cyan correspond to an energy gap smaller than 0.2\,eV. The red region in (b) highlights an unphysical negative energy gap. The calculations are carried out with the PBE functional. A superimposed brown contour line at 0.5\,eV shows the results of published MRCISD calculations\cite{barbatti2014surface}. Comparison with the 0.5\,eV contour calculated here shows that the shape of the energy gap surface in (a), obtained from the broken-spin-symmetry calculation, reproduces the higher-level multireference calculation well near the CoIn.
    }
	\label{fig:fig2}
\end{figure}

Figure \ref{fig:fig2} shows the energy gap between the \added[id=YS]{singly excited V state} and the ground state, N, obtained with the PBE functional as a function of $\theta$ and $\phi$, in order to visualize the CoIn.
% ----------------  figure  2 ----------------------
% -----------------------------------------------
Results for both broken-spin-symmetry and spin-restricted solutions for the N state are shown, while the energy of the V state is always spin-purified (eq \ref{eq:eq5}). For comparison, an isocontour corresponding to a V-N energy gap of 0.5\,eV taken from reported MRCISD calculations\cite{barbatti2014surface} using CAS(2,2) and the aug-cc-pVTZ basis set\cite{Woon1994,Kendall1992,Dunning1989} is shown. There, the CoIn was calculated to be at $\theta = 90^{\circ}$ and $\phi \approx 105^{\circ}$. When the ground state is taken to be the broken-spin-symmetry N solution, the calculated energy gap is below 0.5\,eV only in a small region around $\theta = 90^{\circ}$ and $\phi \approx 100^{\circ}$, where the gap reaches a minimum, in good agreement with the multireference results. On the other hand, when the ground state is taken to be the spin-restricted N solution, the V-N gap falls below 0.5\,eV in a more extended region and an unphysical negative energy gap is observed in the immediate proximity of the CoIn. Therefore, the correct topology for the CoIn is obtained when the spin symmetry is broken in the ground state. TDDFT calculations with the BLYP and the B3LYP functionals\cite{barbatti2014surface}, on the other hand, give too small V-N gaps along $\theta = 90^{\circ}$ for an extended range in $\phi$ and a negative energy gap even when symmetry breaking is allowed. Figure S1 in the ESI shows that time-independent KS-DFT calculations with the BLYP functional also predict the correct CoIn topology when symmetry breaking is allowed in the ground state.

The energy of the spin-mixed determinant, $\ket{ x \spinup y}$, is often used to approximate the energy of the open-shell singlet V state without spin purification. The results of this approach are shown in Figures S2 and S3 of the ESI. When the ground state is taken to be the broken-symmetry N solution, the region in which the V-N gap is below 0.5\,eV becomes more extended, but it remains small and the correct topology of the energy gap surface is still obtained. When the energy gap is evaluated from the spin-restricted N and spin-mixed V states, a negative energy gap region extends across almost all geometries with $\theta = 90^{\circ}$. The latter resembles the results reported by Pradhan {\it et al.}\cite{pradhan2018non}, suggesting that spin purification of the V state and spin symmetry breaking in the ground state were missing there. Although the calculations of Pradhan {\it et al.}\cite{pradhan2018non} included Fermi-Dirac smearing of the orbital occupation numbers, this approach does not significantly alter the shape of the potential energy surface of the N state, as shown in Figure S4 of the ESI.

\begin{figure}
	\centering
	\includegraphics[width = \textwidth]{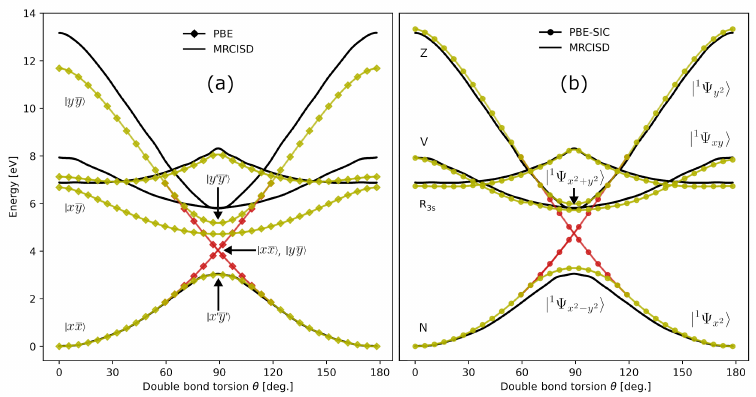}
	\caption{\added[id=YS]{Energy of the N, V and Z states of ethylene as a function of the torsional angle, $\theta$, for $\phi=0$. For completeness, the spin-purified energy of the lowest singlet Rydberg state, R\textsubscript{3s}, is included as well.} The red and yellow curves correspond to the solutions with delocalized and localized frontier orbitals, respectively (see Figure \ref{fig:fig1}). (a) Results of calculations using the PBE functional. The labels refer to the Slater determinants obtained from the time-independent density functional calculations using the PBE or the  self-interaction corrected PBE functional, PBE-SIC. (b) Results of calculations using the PBE-SIC functional. The labels refer to the true, multireference wave functions. For comparison, reported results of MRCISD calculations\cite{Barbatti2004} are shown as black lines in both (a) and (b). For $\theta=0^\circ$, a fully optimized geometry of the molecule in the ground state is obtained for each of the two functionals, PBE and PBE-SIC. 
    }
	\label{fig:fig3}
\end{figure}

\begin{figure}[!h]
    \centering
    \includegraphics[width = 0.95\textwidth]{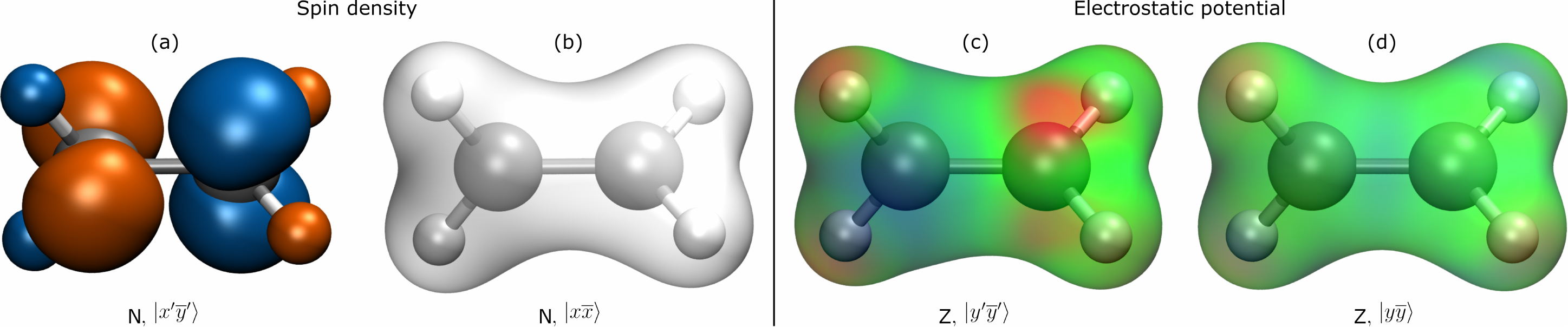}
    \caption{Spin density and electrostatic potential of the N and Z states of $90^{\circ}$ twisted ethylene obtained as single Slater determinants from time-independent density functional calculations using the PBE functional. The isosurfaces are rendered for values of $\pm 0.034$\,a.u. (a) Spin density of the broken-spin-symmetry $\ket{ x^{\prime} \spinup y^{\prime}}$ determinant for the N state. This solution has diradical character. Areas of excess spin-up and spin-down density are shown in orange and blue, respectively. (b) Electron density of the $\ket{ x \spinup x}$ determinant for the N state. This solution has zero spin density everywhere (indicated with a grey isosurface). (c) Electrostatic potential of the $\ket{ y^{\prime} \spinup y^{\prime}}$ determinant for the Z state. This solution has ionic character. Red and blue regions have excess positive and negative charge, respectively. (d) Electrostatic potential of the $\ket{y \spinup y}$ determinant for the Z state. 
    }
    \label{fig:fig4}
\end{figure}

We now analyze the excitation energy of the doubly excited state, Z, along the torsional angle. \added[id=YS]{Figure \ref{fig:fig3} shows the energy along the torsional angle $\theta$ at $\phi=0$ of the N, V and Z states, as well as that of the lowest singlet Rydberg (R\textsubscript{3s}) state, calculated using the PBE and the Perdew-Zunger self-interaction corrected PBE (PBE-SIC) functionals.}
% ----------------  figure 3 ----------------------
% ----------------------------------------------------
For comparison, reported results of MRCI singles and doubles (MRCISD) calculations\cite{Barbatti2004} using a CAS(2,2) and the aux/d-aug-cc-pVDZ\cite{Woon1994,Kendall1992,Dunning1989} basis set are \added[id=YS]{also shown.} At $\theta = 90^\circ$, the spin-restricted determinants with delocalized orbitals, $\ket{ x \spinup x}$ and $\ket{ y \spinup y}$, become degenerate and the energy curves exhibit unphysical cusps, unlike the MRCISD results for the N and Z states, which display an avoided crossing with an energy gap of 2.77\,eV. The cusps at $\theta = 90^\circ$ and the unphysical crossing of the states are highly undesirable as they give rise to incorrect atomic forces and nonadiabatic couplings, the key ingredients of excited state dynamics simulations. \added[id=YS]{We note that variational calculations performed by Mališ and Luber\cite{Malis2020} using a diagonalization-based SCF approach and the hybrid PBE0 functional\cite{adamo1999} also predict a cusp and overestimate the energy of the N state at $\theta = 90^{\circ}$ by ca. 1\,eV. The same authors also report that the employed SCF scheme failed to converge to the Z state for torsional angles larger than 60$^{\circ}$. As can be seen from Figure \ref{fig:fig3}, the broken-symmetry single determinant solutions with localized orbitals, $\ket{ x^{\prime} \spinup y^{\prime}}$ and $\ket{ y^{\prime} \spinup y^{\prime}}$, obtained here with the PBE functional and the direct optimization algorithm} provide energy curves that are in qualitative agreement with the MRCISD results and nearly quantitative agreement is obtained when the PBE-SIC functional is used. The error affecting the spin-restricted determinants with delocalized orbitals results from over- and underestimation of the ionic character of the N and Z states, respectively\cite{Salem1972}, whereas, as illustrated in Figure \ref{fig:fig4}, the broken-symmetry determinants with localized orbitals, $\ket{ x^{\prime} \spinup y^{\prime}}$ and $\ket{ y^{\prime} \spinup y^{\prime}}$, have qualitatively correct diradical and ionic character, respectively, at the perpendicular geometry. We note that the broken-symmetry solutions have finite magnetic and dipole moments, while the multireference wave functions do not. On the other hand, the energy curves are in agreement with the multireference results.
% ----------------  figure 4 ----------------------
% -----------------------------------------------
The considerable improvement in the shape of the curves when the energy is calculated from the broken-symmetry solutions $\ket{ x^{\prime} \spinup y^{\prime}}$ and $\ket{ y^{\prime} \spinup y^{\prime}}$ can be understood by noting that the mixing parameter $\gamma$ becomes 45$^{\circ}$ for $\theta = 90^{\circ}$, as deduced from symmetry considerations (see the ESI). Hence, eqs \ref{eq:eq3} and \ref{eq:eq4} give the (unrelaxed) determinants:
\begin{align}\label{eq:eq6}
    \ket{ x^{\prime} \spinup y^{\prime}} & = \dfrac{1}{\sqrt{2}} \left(\ket{^1\Psi_{x^2-y^2}} - \ket{^3\Psi_{xy}}\right)\,,\\
    \label{eq:eq7}
    \ket{ y^{\prime} \spinup y^{\prime}} & = \dfrac{1}{\sqrt{2}} \left(\ket{^1\Psi_{x^2+y^2}} - \ket{^1\Psi_{xy}}\right) 
\end{align} 
and corresponding energy values:
\begin{align}\label{eq:eq8}
    E(\ket{ x^{\prime} \spinup y^{\prime}}) & = \dfrac{1}{2} \left[E(\ket{^1\Psi_{x^2-y^2}}) + E(\ket{^3\Psi_{xy}})\right] \approx E(\ket{^1\Psi_{x^2-y^2}})\,,\\
    \label{eq:eq9}
    E(\ket{ y^{\prime} \spinup y^{\prime}}) & = \dfrac{1}{2} \left[E(\ket{^1\Psi_{x^2+y^2}}) + E(\ket{^1\Psi_{xy}})\right] \approx E(\ket{^1\Psi_{x^2+y^2}})\,.
\end{align}
The last equality in eq \ref{eq:eq8} follows from the quasi-degeneracy of the ground state singlet and triplet wave functions, $\ket{^1\Psi_{x^2-y^2}}$ and $\ket{^3\Psi_{xy}}$, at the perpendicular geometry, which is well known from previous studies\cite{Salem1972}. The last equality in eq \ref{eq:eq9} follows from the quasi-degeneracy of the singly and doubly excited state wave functions, $\ket{^1\Psi_{xy}}$ and $\ket{^1\Psi_{x^2+y^2}}$, as can be seen in Figure \ref{fig:fig3}. Eqs \ref{eq:eq8} and \ref{eq:eq9} show that, within the approximation of neglecting orbital relaxation, the energy of the determinants $\ket{ x^{\prime} \spinup y^{\prime}}$ and $\ket{ y^{\prime} \spinup y^{\prime}}$ tend to the energy of the multireference wave functions of the N and Z states as the torsional angle is increased towards $90^{\circ}$. The ESI provides an alternative derivation of the same result based on symmetry analysis. 

In the PBE calculations, the excitation energies of the V and Z states from the ground state, N, are underestimated by 0.6 to 1.5 eV, depending on the value of $\theta$. The self-interaction error, which is inherent in practical implementations of KS-DFT, varies for the different states and therefore affects the excitation energy. This argument is supported by Figure S5 in the ESI, which shows the magnitude of the self-interaction correction obtained in the PBE-SIC calculations for the Z and N states at different values of $\theta$. The self-interaction correction for the Z state is always larger than that for the N state, indicating that the underestimation of the Z-N energy gap in the PBE calculations is a consequence of an imbalance in the self-interaction error. As seen in Figure \ref{fig:fig4}, inclusion of the self-interaction correction is needed in order to remove such imbalance and obtain an accurate estimate of the excitation energy for the V and Z states. \added[id=YS]{The excitation energy of the R\textsubscript{3s} state is affected less by the self-interaction error, which is consistent with previous observations,\cite{Ivanov2021jctc} but we note that the ordering of the states at $\theta = 0^{\circ}$ is incorrect without it.} The self-interaction correction also affects the shape of the energy curve for the Z state, as is illustrated in Figure S6 in the ESI. The broken-symmetry Z solution, $\ket{ y^{\prime} \spinup y^{\prime}}$, calculated with PBE underestimates the variation of the energy when going from $\theta = 0^{\circ}$ to $\theta = 90^{\circ}$ by 0.86 eV because the self-interaction error in the Z state is considerably larger at the perpendicular geometry than at the planar geometry (see Figure S5 in the ESI). By using PBE-SIC, close agreement with the MRCISD results is obtained, as shown in Figure \ref{fig:fig3}. The self-interaction correction, therefore, improves both the energy difference between the electronic states and the shape of the energy curve of the Z state.

%  Discussion

In calculations of the dynamics of atoms in excited states, unphysical crossings of different states and regions of a negative energy gap are highly problematic. Typically, a trajectory is generated by advancing the atoms by a time step based on the atomic forces and the electronic structure is calculated at the new geometry by using the occupied orbitals found at the previous step as an initial guess. If this approach is used in calculations of the dynamics in the Z state of ethylene, starting from the planar structure, the calculations converge to the $\ket{y \spinup y}$ Z solution with unphysical energy curve beyond the Coulson-Fischer point and eventually to the spin-restricted $\ket{ x \spinup x}$ N solution after the crossing of the $\ket{y \spinup y}$ and $\ket{ x \spinup x}$ curves, instead of finding the broken-symmetry $\ket{ y^{\prime} \spinup y^{\prime}}$ Z solution that gives the qualitatively correct energy for the doubly excited state. \added[id=YS]{Therefore, when standard molecular dynamics algorithms are used, a solution obtained at a given point in configuration space can depend on the trajectory, i.e. it can be path-dependent.} A different strategy is needed to ensure convergence to a broken-symmetry excited state solution corresponding to the physically meaningful energy surface. It is not possible to simply choose the lower energy solution if two solutions are detected, as is commonly done for the ground state\cite{Vaucher2017}, since there is no minimum energy principle for an excited state. While the broken-symmetry Z solution can be obtained by performing an excitation from the orbitals of the broken-symmetry ground state N solution, $\ket{ x^{\prime} \spinup y^{\prime}}$, at the same geometry, as has been done here, this strategy is impractical for excited state dynamics simulations. We note that the single determinant Z solution for ethylene that gives the correct torsional energy curve, including the broken-symmetry $\ket{ y^{\prime} \spinup y^{\prime}}$ solution around $\theta = 90^{\circ}$, is always a second-order saddle point. The solution $\ket{y \spinup y}$ corresponding to an unphysical energy curve, on the other hand, is a first-order saddle point beyond the Coulson-Fisher point. Therefore, the correct excited state can be obtained by converging to the saddle point of the appropriate order. A minimum mode following method targeting excited states with a specific saddle point order for classical molecular dynamics simulations will be presented in a forthcoming publication.

Clearly, broken-symmetry determinants are not eigenfunctions of quantum mechanical operators such as $\hat{S}^{2}$ or point group symmetry operators, unlike the exact wave function. In the case of $\hat{S}^{2}$, it has been argued\cite{Grafenstein2000,Pople1995} that a single determinant can be allowed to be spin-contaminated in KS-DFT because the KS wave function describes a system of noninteracting electrons, which differs from the true multielectron system. The quality of a broken-symmetry description needs to be assessed with respect to other properties, such as the on-top pair density\cite{Grafenstein2000,Perdew1995}. It remains to be seen how symmetry breaking affects nonadiabatic couplings between electronic states. However, this question might not be of concern for practical applications, insofar as novel mixed quantum-classical algorithms, such as the one presented in ref \cite{Shu2022}, can be used to simulate nonadiabatic dynamics using only the energy and its gradient.

%  Summary

In summary, we have found that the energy of the \added[id=YS]{N, V and Z and R\textsubscript{3s}} states of the ethylene molecule can be obtained over a wide range of the double bond torsion and pyramidalization angles, even in the vicinity of the V-N CoIn and Z-N avoided crossing by using variational, time-independent density functional calculations as long as spin or spatial symmetry are allowed to break. The broken-symmetry solutions arise near electronic degeneracies and coexist with solutions that preserve symmetries but yield energy curves with unphysical crossings. The calculations converge more easily to such solutions with incorrect energy surfaces when the initial guess at a given geometry is constructed from the optimized wave function of the previous step, as is commonly done. This fact could explain why previous variational calculations with the same functional\cite{pradhan2018non} predict an incorrect CoIn topology and negative energy gaps. This problem highlights the need for algorithms that can selectively converge on physically meaningful excited state solutions in nonadibatic dynamics simulations based on time-independent \added[id=YS]{approaches\cite{Vandaele2022, Malis2020, pradhan2018non}.} The application of an explicit Perdew-Zunger self-interaction correction improves the relative energy\added[id=YS]{, the ordering of the states} as well as the shape of the calculated energy curves. The energy gap between the ground and the doubly excited state is underestimated by 0.6 to 1.5\,eV in calculations with the PBE functional, but agrees closely with reported results of multireference calculations when PBE-SIC is used. 

\added[id=YS]{The results presented here indicate that it can be feasible to use a single-determinant variational density functional approach in nonadiabatic excited state dynamics simulations, provided that the calculations are made to converge to appropriate solutions for all states involved in the dynamics up to the initially photoexcited one, and that the functional ensures adequate cancellation of the self-interaction error inherent in practical implementations of KS-DFT.} Since the computational scaling of self-interaction corrected calculations is the same as for calculations with semilocal functionals, this approach can be applied to large systems for which multireference methods are not feasible.
\section*{Computational Methods}
All calculations are performed with the Grid-based Projector Augmented Wave (GPAW) software\cite{GPAW2,GPAW1} with Libxc\cite{LEHTOLA20181} version 4.0.4, using the frozen core approximation and the PAW approach.\cite{PhysRevB.50.17953} Valence electrons are represented by a linear combination of atomic orbitals basis set consisting of primitive Gaussian functions from the aug-cc-pVDZ set\cite{Pritchard2019,Woon1994,Kendall1992,Dunning1989} augmented with a single set of numerical atomic orbitals\cite{Rossi2015,GPAWlcao}. The grid spacing is 0.2 Å, while the dimensions of the simulation cell are according to the default cutoff of the numerical representation of the basis functions\cite{Rossi2015}. The calculations are carried out with the exponential transformation direct optimization method implemented in GPAW\cite{Levi2020jctc,Ivanov2021cpc,Ivanov2021jctc}, using a limited-memory BFGS algorithm with inexact line search for the ground state\cite{Ivanov2021cpc} and a limited-memory SR-1 algorithm with maximum step length of 0.2 for the excited states\cite{Levi2020jctc}. All calculations use real orbitals. The calculations are fully variational, therefore, no orthogonality constraints to lower-energy states are enforced.

For the calculations of the energy surfaces, the geometry of ethylene is first optimized for the ground state using either PBE or PBE-SIC. The energy is subsequently calculated by scanning along the double bond torsion and methylene pyramidalization, while keeping all other internal degrees of freedom fixed. Ground state broken-symmetry solutions are obtained using a linear combination of the symmetric solutions with delocalized orbitals, as given by eqs\,(1) and (2). Broken-symmetry solutions for the doubly excited state are obtained starting from the orbitals of the broken-symmetry ground state at the same geometry by swapping occupation numbers of the HOMO and LUMO in one spin channel and relaxing this initial guess, or using the excited state broken-symmetry solution at another geometry.
\begin{suppinfo}
The authors confirm that the data supporting the findings of this study are available within the article and/or its supplementary materials.

V-N energy gap surfaces from calculations with the BLYP functional; energy surface of the V state estimated from PBE calculations of a mixed-spin determinant and energy gap surface with respect to broken-spin-symmetry and spin-restricted solutions for the ground state, N; torsional energy curve of the spin-restricted solution for the N state obtained from PBE calculations with Fermi-Dirac smearing of the occupation numbers; Perdew-Zunger self-interaction correction for the symmetric and broken-symmetry solutions for the N and Z states as a function of the torsional angle; torsional energy curves from PBE and PBE-SIC calculations adjusted to the maximum energy of the Z state; convergence with respect to grid spacing of the vertical excitation energy of the spin-mixed solution for the V state calculated with PBE; derivation of expressions for the energy of the multireference wave functions as a function of single determinant energies using symmetry analysis.
\end{suppinfo}
\begin{acknowledgement}
This work was supported by the Icelandic Research Fund (grant agreements nos. 217751, 196070, 217734). The calculations were carried out at the Icelandic High Performance Computing Center (IHPC).
\end{acknowledgement}
\bibliography{main_References}
\end{document}

\tableofcontents

\clearpage
\section{BLYP V-N energy gap}
\begin{figure}[!h]
    \centering
    \includegraphics[width = \textwidth]{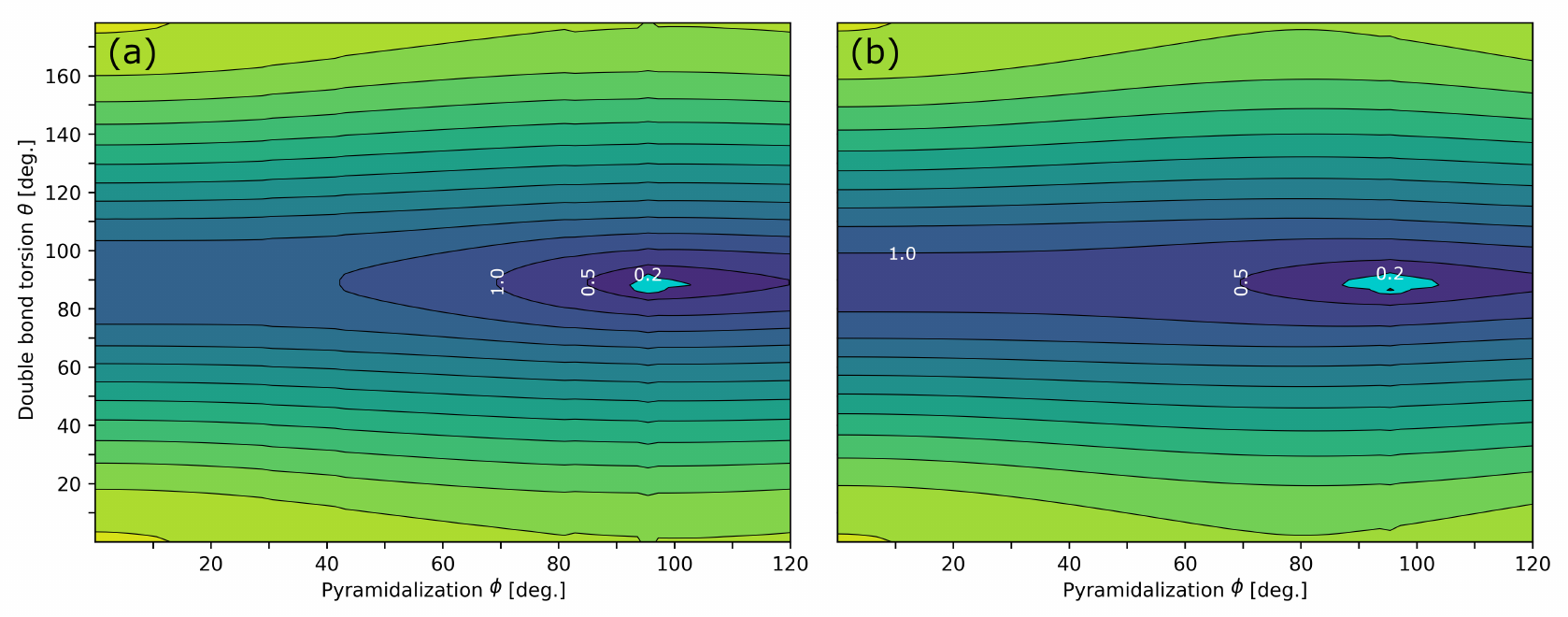}
    \caption{V-N energy gap calculated with the BLYP functional along the double bond torsion and methylene pyramidalization coordinates of the ethylene molecule. In both cases, the ground state, N, is taken to be the broken-spin-symmetry solution $\ket{ x^{\prime} \spinup y^{\prime}}$. The energy of the open-shell singlet excited state, V, is estimated from (a) the spin purification formula (eq\,5 in the main text) using calculations of the spin-mixed determinant $\ket{x \spinup y}$ and triplet determinant $\ket{xy}$, and (b) a calculation of the spin-mixed determinant $\ket{x \spinup y}$. The conical intersection is observed around $\theta = 90^{\circ}$ and $\phi \approx 95^{\circ}$. Regardless of spin purification, no negative energy gap is observed. The region of small energy gap is well-localized if the energy of the spin-mixed V solution is used and it becomes more localized if spin purification is applied.}
    \label{fig:S1}
\end{figure}

\clearpage
\section{PBE energy surfaces with spin-mixed V solution}
\begin{figure}[!h]
    \centering
    \includegraphics[width = \textwidth]{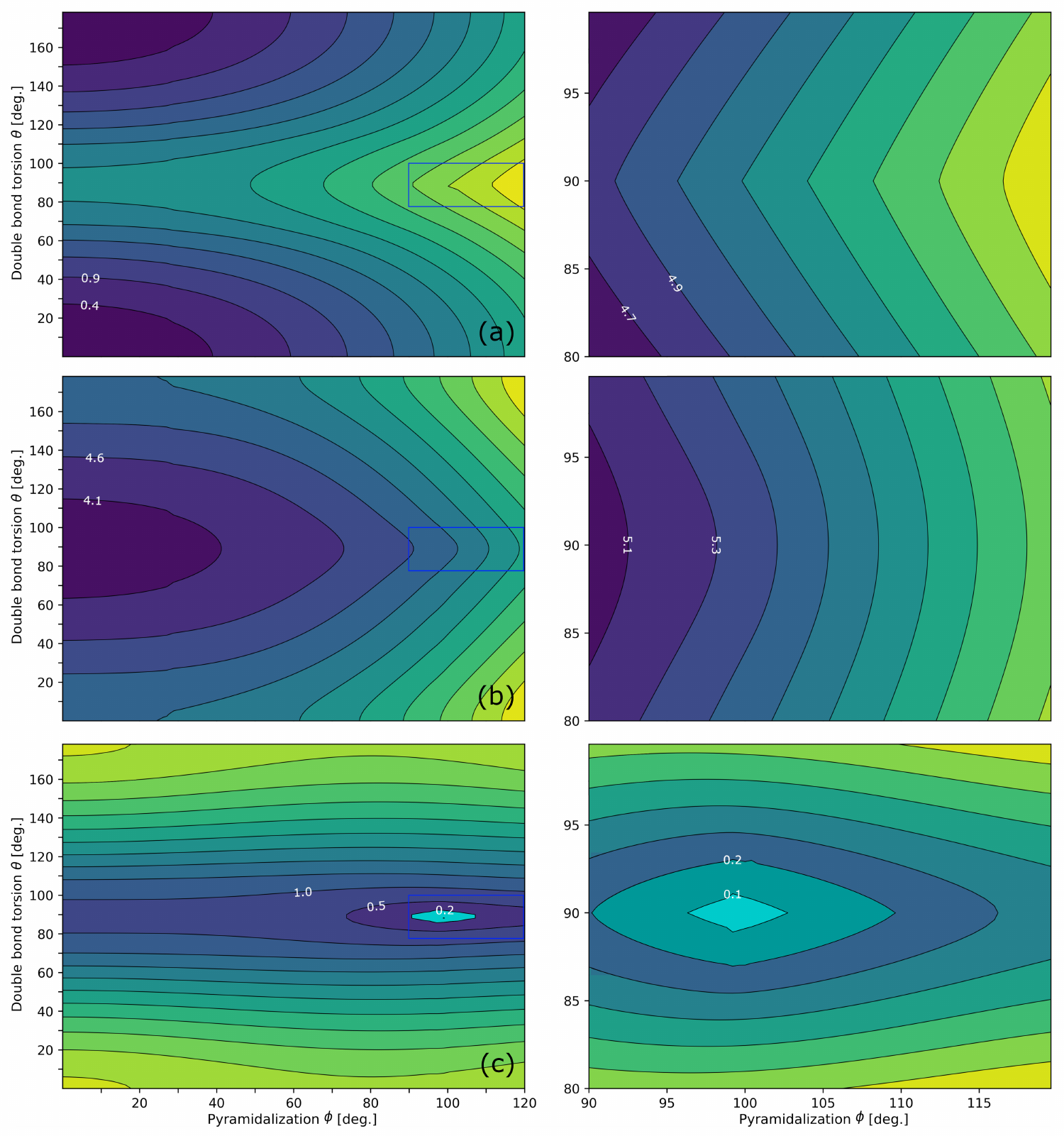}
    \caption{Energy surfaces of the N and V states of ethylene calculated with the PBE functional along the double bond torsion and pyramidalization coordinates of ethylene. (a) Energy of the broken-spin-symmetry N solution, $\ket{ x^{\prime} \spinup y^{\prime}}$. (b) Energy of the spin-mixed V solution,  $\ket{x \spinup y}$. (c) Energy gap between the spin-mixed V and broken-spin-symmetry N solutions. The right column shows a zoom of each surface around the conical intersection.}
    \label{fig:S2}
\end{figure}

\clearpage
\section{PBE energy gap between spin-restricted N and spin-mixed V solutions}
\begin{figure}[!h]
    \centering
    \includegraphics[width = \textwidth]{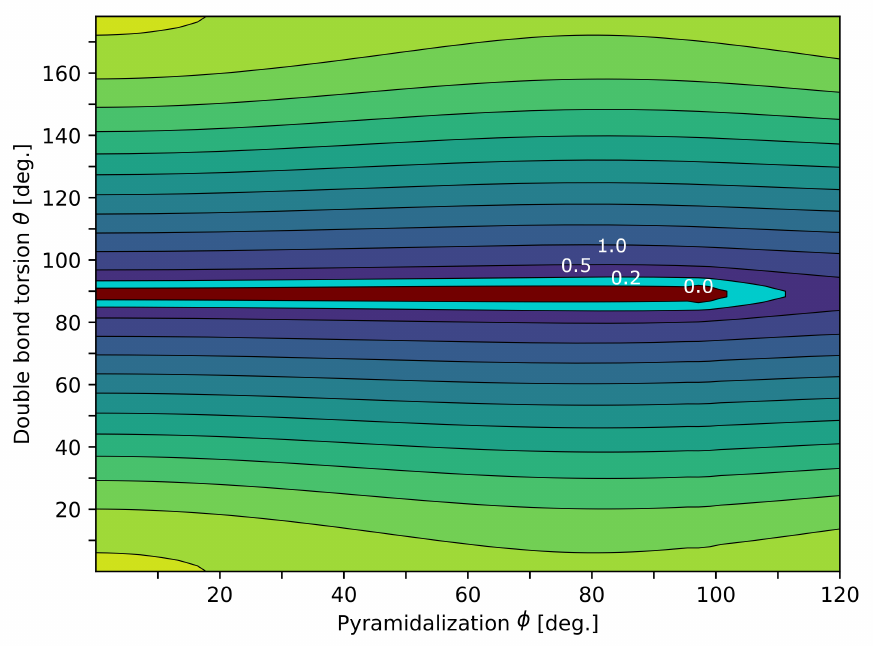}
    \caption{Energy gap between the spin-mixed V solution, $\ket{x \spinup y}$, and the spin-restricted N solution, $\ket{x \spinup x}$, of ethylene obtained with PBE along the double bond torsion and methylene pyramidalization coordinates. Red regions correspond to negative energy gaps.}
    \label{fig:S3}
\end{figure}

\clearpage
\section{PBE ground state torsional energy curve with Fermi-Dirac smearing}
\begin{figure}[!h]
    \centering
    \includegraphics[width = \textwidth]{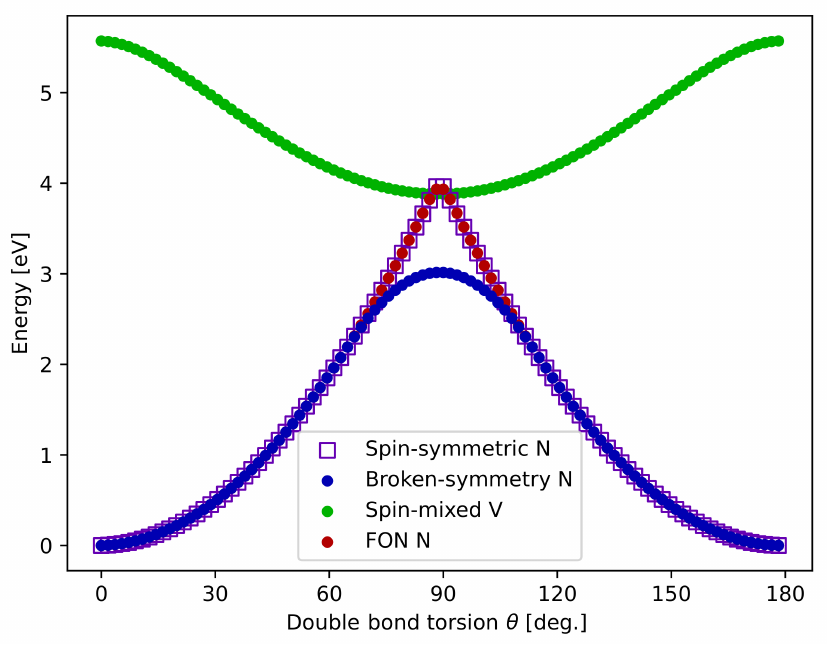}
    \caption{Torsional energy curves of the spin-restricted N solution, $\ket{x \spinup x}$, broken-spin-symmetry N solution, $\ket{ x^{\prime} \spinup y^{\prime}}$, and spin-mixed V solution,  $\ket{x \spinup y}$, of ethylene calculated with PBE at a pyramidalization angle $\phi = 72^{\circ}$. The spin-restricted N solution, $\ket{x \spinup x}$, is calculated with integer occupation numbers, as in Figure S3, and with fractional occupation numbers (FONs) according to a Fermi-Dirac distribution with width of 0.05\,eV (default in GPAW). The energy curve of the spin-restricted N solution with fractional occupation numbers does not differ significantly from the curve of the spin-restricted N solution with integer occupation numbers. Close to $\theta=90^\circ$, the curves of the spin-restricted N solution with integer and fractional occupation numbers are both above the curve of the spin-mixed V solution, creating a negative V-N energy gap. Larger smearing does not make the gap positive.}
    \label{fig:S4}
\end{figure}

\clearpage
\section{Effects of the self-interaction correction on the torsional energy curves for the N and Z solutions}
The energy in a calculation with the Perdew-Zunger self-interaction correction is estimated from
\begin{align}\label{eq:eqS1}
E_\mathrm{SIC}[\{n_{i}\}] = E_\mathrm{KS}[n] - \sum_{i} \left( E_\mathrm{C}[n_{i}] +  E_{xc}[n_{i}, 0]\right)
\end{align}
where $E_\mathrm{KS}$ is the Kohn-Sham energy, $n$ is the total electron density, $n_{i} = |\psi_{i}|^{2}$ is the density of spin orbital $\psi_{i}$, $E_\mathrm{C}$ is the classical Coulomb interaction of a spin orbital with itself, $E_{xc}$ is the self-exchange-correlation energy term, and the sum runs over all occupied spin orbitals. Figure S5 shows the variation of the net self-interaction correction evaluated at the end of a PBE-SIC calculation as $- \sum_{i} \left( E_\mathrm{C}[n_{i}] +  E_{xc}[n_{i}, 0]\right)$ along the torsional angle for the solutions obtained with both delocalized and localized orbitals for the ground state, N, and doubly excited state, Z. 
\begin{figure}[!h]
    \centering
    \includegraphics[width = \textwidth]{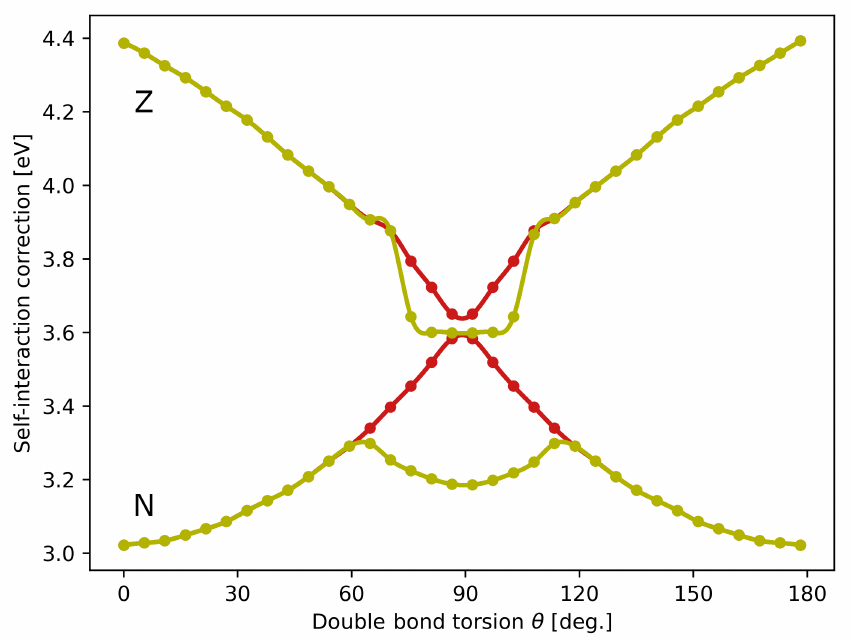}
    \caption{Perdew-Zunger self-interaction correction obtained from variational PBE-SIC calculations of the ground state, N, and lowest doubly excited state, Z, of ethylene along the double bond torsion. The lower red and yellow curves correspond to the spin-restricted and broken-spin-symmetry N solutions, $\ket{x \spinup x}$ and $\ket{ x^{\prime} \spinup y^{\prime}}$, respectively. The upper red and yellow curves correspond to the symmetric and broken-symmetry Z solutions, $\ket{ y \spinup y}$ and $\ket{ y^{\prime} \spinup y^{\prime}}$, respectively.}
    \label{fig:S5}
\end{figure}
From Figure S5, it can be seen that the self-interaction correction for the solutions of the Z state is always larger than the self-interaction correction for the N solution at the ground state geometry. This imbalance is to be expected since the self-exchange-correlation term, which is negative, is usually larger for antibonding orbitals, as those of the Z solutions, than for bonding orbitals of the ground state\cite{Levi2020jctc}. This finding indicates that the self-interaction error in the PBE solutions of the Z state is larger (more negative) than the self-interaction error in the PBE ground state solution, explaining the underestimation of the excitation energy seen in Figure 3 of the main text.  

The main text shows that the spin-restricted solutions with delocalized orbitals $\ket{ x \spinup x}$ and $\ket{ y \spinup y}$, corresponding to approximations of the ground and doubly excited states of ethylene, N and Z, respectively, have qualitatively incorrect torsional energy curves around a torsional angle of $\theta = 90^\circ$ since they display unphysical cusps and crossings. Here, we focus only on the effect of the self-interaction correction on the difference between the energy of the perpendicular ($\theta = 90^\circ$) and planar ($\theta = 0^\circ$) geometries of ethylene for the solutions $\ket{ x \spinup x}$ and $\ket{ y \spinup y}$. For the ground state, N, this energy difference represents the barrier of torsion. As can be seen from Figure 3 in the main text, the energy barrier estimated from the $\ket{ x \spinup x}$ N solution obtained with PBE is closer to the barrier given by MRCISD calculations than the barrier of the $\ket{ x \spinup x}$ N solution calculated with PBE-SIC by $\sim$0.5 eV. The same trend is obtained for the energy difference between perpendicular and planar geometries for the $\ket{ y \spinup y}$ Z solution: The energy difference estimated from the PBE solution is closer to the MRCISD energy difference than for the PBE-SIC solution by $\sim$0.75 eV, as highlighted in Figure S6. From Figure \ref{fig:S5}, it can be seen that the self-interaction correction increases and decreases towards $\theta = 90^\circ$ for $\ket{ x \spinup x}$ and $\ket{ y \spinup y}$, respectively, which suggests that the self-interaction error in PBE at the perpendicular geometry is larger (smaller) than the self-interaction error at the planar geometry for $\ket{ x \spinup x}$ ($\ket{ y \spinup y}$). Static correlation at the perpendicular geometry would also decrease (increase) the energy of the N (Z) state with respect to the planar geometry. However, static correlation is missing in the $\ket{ x \spinup x}$ and $\ket{ y \spinup y}$ solutions. By decreasing and increasing the energy of the $\ket{ x \spinup x}$ and $\ket{ y \spinup y}$ solutions, respectively, at the perpendicular geometry compared to the planar geometry, the self-interaction error in PBE compensates for the lack of static correlation. This compensation of errors can explain why the PBE energy differences between perpendicular and planar geometries for the symmetric solutions $\ket{ x \spinup x}$ and $\ket{ y \spinup y}$ are closer to the MRCISD results than for PBE-SIC. An analogous cancellation of errors occurs in spin-restricted calculations with local and semilocal functionals of the bond stretching in the dihydrogen molecule in the ground state, as noted before\cite{Grafenstein2004, Cremer2002}.

\begin{figure}[!h]
    \centering
    \includegraphics[width = \textwidth]{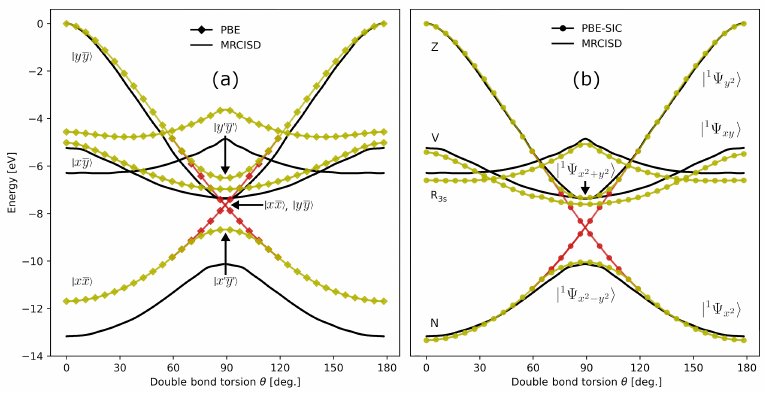}
    \caption{\added[id=YS]{Energy of the N, V and Z states of ethylene as a function of the torsional angle, $\theta$, for a pyramidalization angle $\phi=0$. For completeness, the spin-purified energy of the lowest singlet Rydberg state, R\textsubscript{3s}, is included as well.} The red and yellow curves correspond to symmetric and broken-symmetry solutions obtained for localized and delocalized frontier orbitals, respectively (see Figure 1 in the main text). The zero of the curves was chosen as the maximum of the Z solution at $\theta = 0^\circ$. (a) Results of calculations using the PBE functional. The labels refer to the Slater determinants obtained from the time-independent density functional calculations. (b) Results of calculations using the self-interaction corrected PBE functional, PBE-SIC. The labels refer to the true, multireference wave functions. For comparison, the solid lines in both (a) and (b) show reported results of calculations using the MRCISD method (see main text). For $\theta=0^\circ$, a fully optimized geometry of the molecule in the ground state is obtained for each of the two functionals, PBE and PBE-SIC. }
    \label{fig:S6}
\end{figure}

We now discuss the effect of the self-interaction correction on the broken-symmetry $\ket{ x^{\prime} \spinup y^{\prime}}$ and $\ket{ y^{\prime} \spinup y^{\prime}}$ solutions with delocalized orbitals, which, as shown in the main text, provide qualitatively correct shapes for the torsional energy curves of the ground and doubly excited states, N and Z. Figure 3 in the main text shows that the energy curve of the broken-spin-symmetry solution with localized orbitals $\ket{ x^{\prime} \spinup y^{\prime}}$ for the ground state calculated with the PBE functional approximates the MRCISD curve slightly better than the curve calculated with PBE-SIC, particularly around the Coulson-Fischer points. The torsional barrier is reproduced by PBE within only $\sim$38\,meV. On the other hand, as can be appreciated in Figure \ref{fig:S6}, the broken-symmetry solution $\ket{ y^{\prime} \spinup y^{\prime}}$ for the Z state obtained with PBE underestimates the energy difference between the perpendicular and planar geometries by $\sim$0.86 eV with respect to MRCISD, while PBE-SIC gives an accurate estimate. Figure \ref{fig:S5} shows the net self-interaction correction for $\ket{ x^{\prime} \spinup y^{\prime}}$ and $\ket{ y^{\prime} \spinup y^{\prime}}$ as a function of the torsional angle. The self-interaction correction for $\ket{ x^{\prime} \spinup y^{\prime}}$ reaches a maximum around the Coulson-Fischer points and then decreases towards $\theta = 90^{\circ}$. This behavior suggests that the amount of self-interaction error in the $\ket{ x^{\prime} \spinup y^{\prime}}$ solution lowers the energy of the ground state around the Coulson-Fischer point, leading to an improvement of the energy curve compared to PBE-SIC. For the $\ket{ y^{\prime} \spinup y^{\prime}}$ excited state solution, the self-interaction correction is considerably larger at the perpendicular geometry than at the planar geometry, which indicates that the underestimation of the depth of the Z energy curve in the PBE calculations stems from the combined effect of the self-interaction error and the static correlation introduced through the form of the broken-symmetry wave function, as both raise the energy at the perpendicular geometry.

\clearpage
\section{Torsional energy curves adjusted to the Z state maximum}

\clearpage
\section{Grid convergence}
\begin{figure}[!h]
    \centering
    \includegraphics[width = \textwidth]{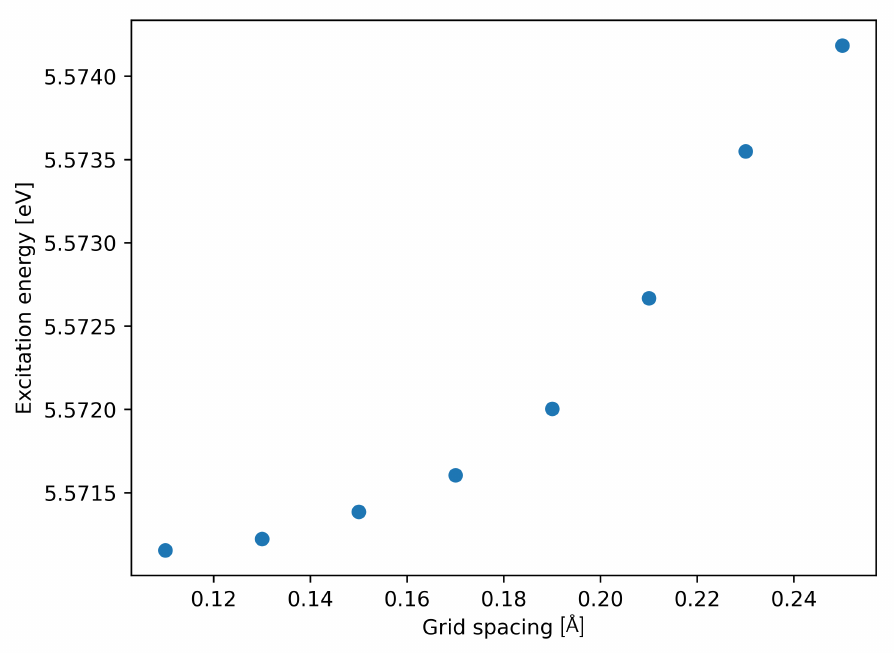}
    \caption{Vertical excitation energy of the spin-mixed V solution, $\ket{x \spinup y}$, of ethylene calculated with PBE at the fully optimized ground state geometry as a function of the grid spacing. At a grid spacing of 0.2\,\AA, the excitation energy is converged below 1\,meV.}
    \label{fig:S7}
\end{figure}

\clearpage
\section{Multideterminant energies from symmetry analysis}
Here, expressions for the energies of the symmetry-adapted multireference wave functions using single determinant energies are derived based on point group symmetry analysis for the 90$^\circ$ twisted geometry of ethylene. Since the lower orbitals are doubly occupied, their irreducible representations square (in a direct product sense) to yield the totally symmetric representation, therefore, it is sufficient to consider only the two frontier orbitals. Orbital relaxation effects are neglected.

The 90$^\circ$ twisted geometry of ethylene has D\textsubscript{2d} point group symmetry. In this point group, there are two possible orbital sets for the HOMO and LUMO. The first set is a localized set consisting of p-type orbitals, whilst the second set is delocalized and comprised of positive and negative linear combinations of the localized orbitals. 

\begin{figure}
    \centering
    \includegraphics[width = 0.4\textwidth]{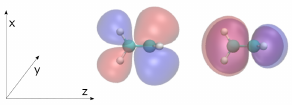}
    \caption{Localized frontier orbital set of ethylene in the point group D\textsubscript{2d} with isovalues of $\pm 0.034$\,a.u.}
    \label{fig:d2d_localized}
\end{figure}

Figure \ref{fig:d2d_localized} shows the localized set of degenerate orbitals. This set of frontier orbitals, denoted as $x'$, $y'$, has considerable p-orbital character. Table \ref{tab:d2d_characters_localized} collects the symmetry characters of this orbital set.

\begin{table}
    \centering
    \caption{Symmetry characters of the localized frontier orbital set of ethylene in the point group D\textsubscript{2d}.}
    \begin{tabular}{|c|c|c|c|c|c|c|c|c|}
        \hline
        \  & E & 2S\textsubscript{4} & C\textsubscript{2}(z) & 2C\textsubscript{2}' & 2$\sigma$\textsubscript{d} & \ \\
        \hline
        $\left(x', y'\right)$ & $\begin{pmatrix}1 & 0\\0 & 1\end{pmatrix}$ & $\pm\begin{pmatrix}0 & 1\\-1 & 0\end{pmatrix}$ & $\begin{pmatrix}-1 & 0\\0 & -1\end{pmatrix}$ & $\pm\begin{pmatrix}0 & 1\\1 & 0\end{pmatrix}$ &  $\pm\begin{pmatrix}1 & 0\\0 & -1\end{pmatrix}$ & e\\
        \hline
    \end{tabular}
    \label{tab:d2d_characters_localized}
\end{table}

\begin{figure}
    \centering
    \includegraphics[width = 0.4\textwidth]{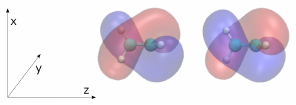}
    \caption{Delocalized frontier orbital set of ethylene in the point group D\textsubscript{2d} with isovalues of $\pm 0.034$\,a.u.}
    \label{fig:d2d_delocalized}
\end{figure}

The delocalized orbital set is visualized in fig.\,\ref{fig:d2d_delocalized}. This set is obtained from the localized set by positive and negative linear combination as

\begin{align*}
    x & = \frac{1}{\sqrt{2}}\left(x' + y'\right)\,,\\
    y & = \frac{1}{\sqrt{2}}\left(x' - y'\right)\,.
\end{align*}

While the $\pi$ nodal planes of the localized set are aligned normal to the x- and y-axes, the $\pi$ nodal planes of the delocalized set are curved. Its symmetry characters are gathered in tab. \ref{tab:d2d_characters_delocalized}.

\begin{table}
    \centering
    \caption{Symmetry characters of the delocalized frontier orbital set of ethylene in the point group D\textsubscript{2d}.}
    \begin{tabular}{|c|c|c|c|c|c|c|c|c|}
        \hline
        \  & E & 2S\textsubscript{4} & C\textsubscript{2}(z) & 2C\textsubscript{2}' & 2$\sigma$\textsubscript{d} & \ \\
        \hline
        $\left(x, y\right)$ & $\begin{pmatrix}1 & 0\\0 & 1\end{pmatrix}$ & $\pm\begin{pmatrix}0 & 1\\-1 & 0\end{pmatrix}$ & $\begin{pmatrix}-1 & 0\\0 & -1\end{pmatrix}$ & $\pm\begin{pmatrix}1 & 0\\0 & -1\end{pmatrix}$ &  $\pm\begin{pmatrix}0 & 1\\1 & 0\end{pmatrix}$ & e\\
        \hline
    \end{tabular}
    \label{tab:d2d_characters_delocalized}
\end{table}

Both sets belong to the irreducible representation e. The characteristic symmetry difference between the two sets manifests most notably in their characters with respect to rotation around the secondary symmetry axes and reflection at the dihedral symmetry planes, for which their characters are swapped.

There are six antisymmetrized determinants which can be formed with the localized orbital set:

\begin{align*}
D'_1 & = \ket{x'_{1}\alpha_{1}x'_{2}\beta_{2}}\,,\\
D'_2 & = \ket{y'_{1}\alpha_{1}y'_{2}\beta_{2}}\,,\\
D'_3 & = \ket{x'_{1}\alpha_{1}y'_{2}\beta_{2}}\,,\\
D'_4 & = \ket{y'_{1}\alpha_{1}x'_{2}\beta_{2}}\,,\\
D'_5 & = \ket{x'_{1}\alpha_{1}y'_{2}\alpha_{2}}\,,\\
D'_6 & = \ket{x'_{1}\beta_{1}y'_{2}\beta_{2}}\,.
\end{align*}

The first two determinants are singlet determinants, the second pair consists of spin impure determinants and the last two determinants are triplet determinants. The characters of these determinants are displayed in tab.\,\ref{tab:d2d_characters_determinants}. The spatial parts of $D'_{5}$ and $D'_{6}$ share the symmetry character of $D'_{3}$.

\begin{table}
    \centering
    \caption{Symmetry characters of the determinants in the point group D\textsubscript{2d}.}
    \begin{tabular}{|c|c|c|c|c|c|}
    \hline
        \  & E & 2S\textsubscript{4} & C\textsubscript{2}(z) & 2C\textsubscript{2}' & 2$\sigma$\textsubscript{d}\\
        \hline
        $\left(D'_{1}, D'_{2}\right)$ & $\begin{pmatrix}1 & 0\\0 & 1\end{pmatrix}$ & $\begin{pmatrix}0 & 1\\1 & 0\end{pmatrix}$ & $\begin{pmatrix}1 & 0\\0 & 1\end{pmatrix}$ & $\begin{pmatrix}0 & 1\\1 & 0\end{pmatrix}$ &  $\begin{pmatrix}1 & 0\\0 & 1\end{pmatrix}$\\
        \hline
        $\left(D'_{3}, D'_{4}\right)$ & $\begin{pmatrix}1 & 0\\0 & 1\end{pmatrix}$ & $\begin{pmatrix}0 & -1\\-1 & 0\end{pmatrix}$ & $\begin{pmatrix}1 & 0\\0 & 1\end{pmatrix}$ & $\begin{pmatrix}0 & 1\\1 & 0\end{pmatrix}$ &  $\begin{pmatrix}-1 & 0\\0 & -1\end{pmatrix}$\\
        \hline
    \end{tabular}
    \label{tab:d2d_characters_determinants}
\end{table}

The spin parts between different singlet and spin impure determinants integrate to unity, so a spatial symmetry analysis is required to show which off-diagonal Hamiltonian matrix elements in the set of singlet and spin impure determinants vanish.

We see that for the first two determinants,

\begin{align*}
    \hat{E}\bra{D'_{1}}\hat{H}\ket{D'_{2}} & = \bra{D'_{1}}\hat{H}\ket{D'_{2}}\,,\\
    \hat{S}_{4}\bra{D'_{1}}\hat{H}\ket{D'_{2}} & = \bra{D'_{2}}\hat{H}\ket{D'_{1}} = \bra{D'_{1}}\hat{H}\ket{D'_{2}}\,.
\end{align*}
Since all other symmetry operations behave as $\hat{E}$ or $\hat{S}_{4}$ it follows that
\begin{equation*}
    \bra{D'_{1}}\hat{H}\ket{D'_{2}}\neq 0\,.
\end{equation*}

For the last two singlet determinants,

\begin{align*}
    \hat{E}\bra{D'_{3}}\hat{H}\ket{D'_{4}} & = \bra{D'_{3}}\hat{H}\ket{D'_{4}}\,,\\
    \hat{S}_{4}\bra{D'_{3}}\hat{H}\ket{D'_{4}} & = \bra{\left(-D'_{4}\right)}\hat{H}\ket{\left(-D'_{3}\right)} = \bra{D'_{4}}\hat{H}\ket{D'_{3}} = \bra{D'_{3}}\hat{H}\ket{D'_{4}}\,.\\
\end{align*}
Since all other symmetry operations behave as $\hat{E}$ or $\hat{S}_{4}$ it follows that
\begin{equation*}
    \bra{D'_{3}}\hat{H}\ket{D'_{4}}\neq 0\,.
\end{equation*}

For $D'_{1}$ and $D'_{4}$,

\begin{align*}
    \hat{\sigma}_{d}\bra{D'_{1}}\hat{H}\ket{D'_{4}} & = \bra{D'_{1}}\hat{H}\ket{\left(-D'_{4}\right)} = -\bra{D'_{1}}\hat{H}\ket{D'_{4}}\,,\\
    \hat{S}_{4}\bra{D'_{1}}\hat{H}\ket{D'_{4}} & = \bra{D'_{2}}\hat{H}\ket{\left(-D'_{3}\right)} = -\bra{D'_{2}}\hat{H}\ket{D'_{3}}\,.
\end{align*}

Since the integrals must adhere to the point group symmetry of the system it follows that

\begin{align*}
    \bra{D'_{1}}\hat{H}\ket{D'_{4}} & = -\bra{D'_{1}}\hat{H}\ket{D'_{4}} = \bra{D'_{2}}\hat{H}\ket{D'_{3}} = -\bra{D'_{2}}\hat{H}\ket{D'_{3}} = 0\,.
\end{align*}

For $D'_{1}$ and $D'_{3}$,

\begin{align*}
    \hat{\sigma}_{d}\bra{D'_{1}}\hat{H}\ket{D'_{3}} & = \bra{D'_{1}}\hat{H}\ket{\left(-D'_{3}\right)} = -\bra{D'_{1}}\hat{H}\ket{D'_{3}}\,,\\
    \hat{S}_{4}\bra{D'_{1}}\hat{H}\ket{D'_{3}} & = \bra{D'_{2}}\hat{H}\ket{\left(-D'_{4}\right)} = -\bra{D'_{2}}\hat{H}\ket{D'_{4}}\,.
\end{align*}

Since the integrals must adhere to the point group symmetry of the system it follows that

\begin{align*}
    \bra{D'_{1}}\hat{H}\ket{D'_{3}} & = -\bra{D'_{1}}\hat{H}\ket{D'_{3}} = \bra{D'_{2}}\hat{H}\ket{D'_{4}} = -\bra{D'_{2}}\hat{H}\ket{D'_{4}} = 0\,.
\end{align*}

The determinants $D'_{1}$ and $D'_{2}$ as well as $D'_{3}$ and $D'_{4}$ are degenerate,

\begin{align*}
    \bra{D'_{1}}\hat{H}\ket{D'_{1}} & = \bra{D'_{2}}\hat{H}\ket{D'_{2}}\,,\\
    \bra{D'_{3}}\hat{H}\ket{D'_{3}} & = \bra{D'_{4}}\hat{H}\ket{D'_{4}}\,.
\end{align*}

Similar determinants can be formed for the delocalized orbital set and are related to the localized set:

\begin{align*}
D_1 & = \ket{x_1\alpha_{1}x_2\beta_{2}} = \frac{1}{2}\left(D'_{1} + D'_{2} + D'_{3} + D'_{4}\right)\,,\\
D_2 & = \ket{y_1\alpha_{1}y_2\beta_{2}} = \frac{1}{2}\left(D'_{1} + D'_{2} - D'_{3} - D'_{4}\right)\,,\\
D_3 & = \ket{x_1\alpha_{1}y_2\beta_{2}} = \frac{1}{2}\left(D'_{1} - D'_{2} - D'_{3} + D'_{4}\right)\,,\\
D_4 & = \ket{y_1\alpha_{1}x_2\beta_{2}} = \frac{1}{2}\left(D'_{1} - D'_{2} + D'_{3} - D'_{4}\right)\,,\\
D_5 & = \ket{x_1\alpha_{1}y_2\alpha_{2}} = -D'_{5}\,,\\
D_6 & = \ket{x_1\beta_{1}y_2\beta_{2}} = -D'_{6}\,.
\end{align*}

The triplet determinants are identical in the localized and delocalized sets and $D_{5}$ and $D_{6}$ can therefore be omitted in the further analysis.

Making use of the fact that the squared spin parts of the determinants each integrate to unity, we can show the relationships between the energies of the determinants using the two orbital sets.

We obtain for the first delocalized determinant energy

\begin{align*}
    \bra{D_{1}}\hat{H}\ket{D_{1}} & = \frac{1}{4}\bra{\left(D'_{1} + D'_{2} + D'_{3} + D'_{4}\right)}\hat{H}\ket{\left(D'_{1} + D'_{2} + D'_{3} + D'_{4}\right)}\\
    & = \frac{1}{4}\left(\bra{D'_{1}}\hat{H}\ket{D'_{1}} + \bra{D'_{1}}\hat{H}\ket{D'_{2}} + \bra{D'_{1}}\hat{H}\ket{D'_{3}} + \bra{D'_{1}}\hat{H}\ket{D'_{4}}\right.\\
    & + \bra{D'_{2}}\hat{H}\ket{D'_{1}} + \bra{D'_{2}}\hat{H}\ket{D'_{2}} + \bra{D'_{2}}\hat{H}\ket{D'_{3}} + \bra{D'_{2}}\hat{H}\ket{D'_{4}}\\
    & + \bra{D'_{3}}\hat{H}\ket{D'_{1}} + \bra{D'_{3}}\hat{H}\ket{D'_{2}} + \bra{D'_{3}}\hat{H}\ket{D'_{3}} + \bra{D'_{3}}\hat{H}\ket{D'_{4}}\\
    & + \left.\bra{D'_{4}}\hat{H}\ket{D'_{1}} + \bra{D'_{4}}\hat{H}\ket{D'_{2}} + \bra{D'_{4}}\hat{H}\ket{D'_{3}} + \bra{D'_{4}}\hat{H}\ket{D'_{4}}\right)\\
    & = \frac{1}{2}\left(\bra{D'_{1}}\hat{H}\ket{D'_{1}} + \bra{D'_{3}}\hat{H}\ket{D'_{3}} + \bra{D'_{1}}\hat{H}\ket{D'_{2}} + \bra{D'_{3}}\hat{H}\ket{D'_{4}}\right)\,.
\end{align*}

We obtain for the second delocalized determinant energy

\begin{align*}
    \bra{D_{2}}\hat{H}\ket{D_{2}}
    & = \frac{1}{4}\bra{\left(D'_{1} + D'_{2} - D'_{3} - D'_{4}\right)}\hat{H}\ket{\left(D'_{1} + D'_{2} - D'_{3} - D'_{4}\right)}\\
    & = \frac{1}{4}\left(\bra{D'_{1}}\hat{H}\ket{D'_{1}} + \bra{D'_{1}}\hat{H}\ket{D'_{2}} - \bra{D'_{1}}\hat{H}\ket{D'_{3}} - \bra{D'_{1}}\hat{H}\ket{D'_{4}}\right.\\
    & + \bra{D'_{2}}\hat{H}\ket{D'_{1}} + \bra{D'_{2}}\hat{H}\ket{D'_{2}} - \bra{D'_{2}}\hat{H}\ket{D'_{3}} - \bra{D'_{2}}\hat{H}\ket{D'_{4}}\\
    & - \bra{D'_{3}}\hat{H}\ket{D'_{1}} - \bra{D'_{3}}\hat{H}\ket{D'_{2}} + \bra{D'_{3}}\hat{H}\ket{D'_{3}} + \bra{D'_{3}}\hat{H}\ket{D'_{4}}\\
    & - \left.\bra{D'_{4}}\hat{H}\ket{D'_{1}} - \bra{D'_{4}}\hat{H}\ket{D'_{2}} + \bra{D'_{4}}\hat{H}\ket{D'_{3}} + \bra{D'_{4}}\hat{H}\ket{D'_{4}}\right)\\
    & = \frac{1}{2}\left(\bra{D'_{1}}\hat{H}\ket{D'_{1}} + \bra{D'_{3}}\hat{H}\ket{D'_{3}} + \bra{D'_{1}}\hat{H}\ket{D'_{2}} + \bra{D'_{3}}\hat{H}\ket{D'_{4}}\right)\\
    & = \bra{D_{1}}\hat{H}\ket{D_{1}}\,.
\end{align*}

We obtain for the third delocalized determinant energy

\begin{align*}
    \bra{D_{3}}\hat{H}\ket{D_{3}}
    & = \frac{1}{4}\bra{\left(D'_{1} - D'_{2} - D'_{3} + D'_{4}\right)}\hat{H}\ket{\left(D'_{1} - D'_{2} - D'_{3} + D'_{4}\right)}\\
    & = \frac{1}{4}\left(\bra{D'_{1}}\hat{H}\ket{D'_{1}} - \bra{D'_{1}}\hat{H}\ket{D'_{2}} - \bra{D'_{1}}\hat{H}\ket{D'_{3}} + \bra{D'_{1}}\hat{H}\ket{D'_{4}}\right.\\
    & - \bra{D'_{2}}\hat{H}\ket{D'_{1}} + \bra{D'_{2}}\hat{H}\ket{D'_{2}} + \bra{D'_{2}}\hat{H}\ket{D'_{3}} - \bra{D'_{2}}\hat{H}\ket{D'_{4}}\\
    & - \bra{D'_{3}}\hat{H}\ket{D'_{1}} + \bra{D'_{3}}\hat{H}\ket{D'_{2}} + \bra{D'_{3}}\hat{H}\ket{D'_{3}} - \bra{D'_{3}}\hat{H}\ket{D'_{4}}\\
    & + \left.\bra{D'_{4}}\hat{H}\ket{D'_{1}} - \bra{D'_{4}}\hat{H}\ket{D'_{2}} - \bra{D'_{4}}\hat{H}\ket{D'_{3}} + \bra{D'_{4}}\hat{H}\ket{D'_{4}}\right)\\
    & = \frac{1}{2}\left(\bra{D'_{1}}\hat{H}\ket{D'_{1}} + \bra{D'_{3}}\hat{H}\ket{D'_{3}} - \bra{D'_{1}}\hat{H}\ket{D'_{2}} - \bra{D'_{3}}\hat{H}\ket{D'_{4}}\right)\,.
\end{align*}

We obtain for the fourth delocalized determinant energy

\begin{align*}
    \bra{D_{4}}\hat{H}\ket{D_{4}}
    & = \frac{1}{4}\bra{\left(D'_{1} - D'_{2} + D'_{3} - D'_{4}\right)}\hat{H}\ket{\left(D'_{1} - D'_{2} + D'_{3} - D'_{4}\right)}\\
    & = \frac{1}{4}\left(\bra{D'_{1}}\hat{H}\ket{D'_{1}} - \bra{D'_{1}}\hat{H}\ket{D'_{2}} + \bra{D'_{1}}\hat{H}\ket{D'_{3}} - \bra{D'_{1}}\hat{H}\ket{D'_{4}}\right.\\
    & - \bra{D'_{2}}\hat{H}\ket{D'_{1}} + \bra{D'_{2}}\hat{H}\ket{D'_{2}} - \bra{D'_{2}}\hat{H}\ket{D'_{3}} + \bra{D'_{2}}\hat{H}\ket{D'_{4}}\\
    & + \bra{D'_{3}}\hat{H}\ket{D'_{1}} - \bra{D'_{3}}\hat{H}\ket{D'_{2}} + \bra{D'_{3}}\hat{H}\ket{D'_{3}} - \bra{D'_{3}}\hat{H}\ket{D'_{4}}\\
    & - \left.\bra{D'_{4}}\hat{H}\ket{D'_{1}} + \bra{D'_{4}}\hat{H}\ket{D'_{2}} - \bra{D'_{4}}\hat{H}\ket{D'_{3}} + \bra{D'_{4}}\hat{H}\ket{D'_{4}}\right)\\
    & = \frac{1}{2}\left(\bra{D'_{1}}\hat{H}\ket{D'_{1}} + \bra{D'_{3}}\hat{H}\ket{D'_{3}} - \bra{D'_{1}}\hat{H}\ket{D'_{2}} - \bra{D'_{3}}\hat{H}\ket{D'_{4}}\right)\\
    & = \bra{D_{3}}\hat{H}\ket{D_{3}}\,.
\end{align*}

Note that $D_{1}$ and $D_{2}$, as well as $D_{3}$ and $D_{4}$ are degenerate. If the difference of the non-degenerate delocalized determinants is taken the diagonal part of the Hamiltonian in the localized representation cancels:

\begin{align*}
    \bra{D_{1}}\hat{H}\ket{D_{1}} - \bra{D_{3}}\hat{H}\ket{D_{3}} & = \bra{D'_{1}}\hat{H}\ket{D'_{2}} + \bra{D'_{3}}\hat{H}\ket{D'_{4}}\,.
\end{align*}

The multideterminant wave functions in terms of the localized orbitals are

\begin{align*}
    \Psi_{\text{T}} & = \frac{1}{\sqrt{2}}\left(x'_{1}y'_{2} - y'_{1}x'_{2}\right)\times \left\{\begin{array}{lll} \alpha_{1}\alpha_{2}\\ \frac{1}{\sqrt{2}}\left(\alpha_{1}\beta_{2} + \beta_{1}\alpha_{2}\right)\\ \beta_{1}\beta_{2}\end{array}\right. = \left\{\begin{array}{lll} D'_{5}\\ \frac{1}{\sqrt{2}}\left(D'_{3} - D'_{4}\right)\\
    D'_{6}\end{array}\right.\,,\\
    \Psi_{\text{N}} & = \frac{1}{2}\left(x'_{1}y'_{2} + y'_{1}x'_{2}\right)\left(\alpha_{1}\beta_{2} - \beta_{1}\alpha_{2}\right) = \frac{1}{\sqrt{2}}\left(D'_{3} + D'_{4}\right)\,,\\
    \Psi_{\text{V}} & = \frac{1}{2}\left(x'_{1}x'_{2} - y'_{1}y'_{2}\right)\left(\alpha_{1}\beta_{2} - \beta_{1}\alpha_{2}\right) = \frac{1}{\sqrt{2}}\left(D'_{1} - D'_{2}\right)\,,\\
    \Psi_{\text{Z}} & = \frac{1}{2}\left(x'_{1}x'_{2} + y'_{1}y'_{2}\right)\left(\alpha_{1}\beta_{2} - \beta_{1}\alpha_{2}\right) = \frac{1}{\sqrt{2}}\left(D'_{1} + D'_{2}\right)\,.
\end{align*}

\begin{table}
    \centering
    \caption{Symmetry characters of the wave functions in the point group D\textsubscript{2d}.}
    \begin{tabular}{|c|c|c|c|c|c|c|}
        \hline
        \  & E & 2S\textsubscript{4} & C\textsubscript{2}(z) & 2C\textsubscript{2}' & 2$\sigma$\textsubscript{d} & \ \\
        \hline
        $\Psi_{\text{T}}$ & 1 & 1 & 1 & -1 & -1 & A\textsubscript{2}\\
        \hline
        $\Psi_{\text{Z}}$ & 1 & 1 & 1 & 1 & 1 & A\textsubscript{1}\\
        \hline
        $\Psi_{\text{N}}$ & 1 & -1 & 1 & -1 & 1 & B\textsubscript{2}\\
        \hline
        $\Psi_{\text{V}}$ & 1 & -1 & 1 & 1 & -1 & B\textsubscript{1}\\
        \hline
    \end{tabular}
    \label{tab:d2d_characters_wave_functions}
\end{table}

Their characters and irreducible representations are shown in tab.\,\ref{tab:d2d_characters_wave_functions}. The following solvable set of equations is obtained:

\begin{align}
    \bra{D'_{5}}\hat{H}\ket{D'_{5}} & = \bra{\Psi_{\text{T}}}\hat{H}\ket{\Psi_{\text{T}}}\,,\label{eq:d2d_1}\\ 
    2\bra{D'_{3}}\hat{H}\ket{D'_{3}} & = \bra{\Psi_{\text{T}}}\hat{H}\ket{\Psi_{\text{T}}} + \bra{\Psi_{\text{N}}}\hat{H}\ket{\Psi_{\text{N}}}\,,\label{eq:d2d_2}\\
    2\bra{D'_{1}}\hat{H}\ket{D'_{1}} & = \bra{\Psi_{\text{Z}}}\hat{H}\ket{\Psi_{\text{Z}}} + \bra{\Psi_{\text{V}}}\hat{H}\ket{\Psi_{\text{V}}}\,,\label{eq:d2d_3}\\
    2\bra{D_{1}}\hat{H}\ket{D_{1}} -2\bra{D_{3}}\hat{H}\ket{D_{3}} & = 
    \bra{\Psi_{\text{N}}}\hat{H}\ket{\Psi_{\text{N}}} - \bra{\Psi_{\text{T}}}\hat{H}\ket{\Psi_{\text{T}}}\label{eq:d2d_4}\\
    & + \bra{\Psi_{\text{Z}}}\hat{H}\ket{\Psi_{\text{Z}}} - \bra{\Psi_{\text{V}}}\hat{H}\ket{\Psi_{\text{V}}}\,.\nonumber
\end{align}

Substituting (\ref{eq:d2d_1}) into (\ref{eq:d2d_2}) yields

\begin{equation}\label{eq:d2d_5}
    \bra{\Psi_{\text{N}}}\hat{H}\ket{\Psi_{\text{N}}} = 2\bra{D'_{3}}\hat{H}\ket{D'_{3}} - \bra{D'_{5}}\hat{H}\ket{D'_{5}}\,.
\end{equation}

Adding and subtracting (\ref{eq:d2d_3}) and (\ref{eq:d2d_4}) yields

\begin{align}
    & 2\bra{D'_{1}}\hat{H}\ket{D'_{1}} + 2\bra{D_{1}}\hat{H}\ket{D_{1}} -2\bra{D_{3}}\hat{H}\ket{D_{3}}\label{eq:d2d_6}\\
    =\ & 2\bra{\Psi_{\text{Z}}}\hat{H}\ket{\Psi_{\text{Z}}} + \bra{\Psi_{\text{N}}}\hat{H}\ket{\Psi_{\text{N}}} - \bra{\Psi_{\text{T}}}\hat{H}\ket{\Psi_{\text{T}}}\,,\nonumber\\
    & 2\bra{D'_{1}}\hat{H}\ket{D'_{1}} - 2\bra{D_{1}}\hat{H}\ket{D_{1}} + 2\bra{D_{3}}\hat{H}\ket{D{3}}\label{eq:d2d_7}\\
    =\ & 2\bra{\Psi_{\text{V}}}\hat{H}\ket{\Psi_{\text{V}}} - \bra{\Psi_{\text{N}}}\hat{H}\ket{\Psi_{\text{N}}} + \bra{\Psi_{\text{T}}}\hat{H}\ket{\Psi_{\text{T}}}\,.\nonumber
\end{align}

Substituting (\ref{eq:d2d_1}) and (\ref{eq:d2d_5}) into (\ref{eq:d2d_6}) and (\ref{eq:d2d_7}) yields

\begin{align*}
    \bra{\Psi_{\text{Z}}}\hat{H}\ket{\Psi_{\text{Z}}} & = \bra{D'_{1}}\hat{H}\ket{D'_{1}} + \bra{D_{1}}\hat{H}\ket{D_{1}} - \bra{D_{3}}\hat{H}\ket{D_{3}} - \bra{D'_{3}}\hat{H}\ket{D'_{3}}\\
    & + \bra{D'_{5}}\hat{H}\ket{D'_{5}}\,,\\
    \bra{\Psi_{\text{V}}}\hat{H}\ket{\Psi_{\text{V}}} & = \bra{D'_{1}}\hat{H}\ket{D'_{1}} - \bra{D_{1}}\hat{H}\ket{D_{1}} + \bra{D_{3}}\hat{H}\ket{D_{3}} + \bra{D'_{3}}\hat{H}\ket{D'_{3}}\\
    & - \bra{D'_{5}}\hat{H}\ket{D'_{5}}\,.
\end{align*}

Note that the energies of $\Psi_{\text{Z}}$ and $\Psi_{\text{V}}$ are equally split around the energy of $D'_{1}$ by
\begin{equation*}
    \Delta E_{1} = \pm \left(\bra{D_{1}}\hat{H}\ket{D_{1}} - \bra{D_{3}}\hat{H}\ket{D_{3}} - \bra{D'_{3}}\hat{H}\ket{D'_{3}} + \bra{D'_{5}}\hat{H}\ket{D'_{5}}\right)\,.
\end{equation*}

As this splitting is very small in the D\textsubscript{2d} point group, it can be concluded that the single determinant energy approximates the energies of both wave functions well, so that
\begin{equation*}
    \bra{\Psi_{\text{Z}}}\hat{H}\ket{\Psi_{\text{Z}}} \approx \bra{\Psi_{\text{V}}}\hat{H}\ket{\Psi_{\text{V}}} \approx \bra{D'_{1}}\hat{H}\ket{D'_{1}}\,.
\end{equation*}

\bibliography{si_References}